\documentclass[10pt,preprint]{aastex}

\usepackage[lowtilde]{url}

\newcommand\etal{et~al.}
\newcommand {\kms}    {km~s$^{-1}$}

\begin{document}

\title{Gaia-EDR3 Parallax Distances to the Great Carina Nebula  \\
and its Star Clusters (Trumpler 14, 15, 16) }     

\author{ J. Michael Shull, Jeremy Darling, and Charles Danforth \\
Department of Astrophysical and Planetary Sciences and CASA \\
University of Colorado, 389-UCB, Boulder, CO 80309 
}

\email{michael.shull@colorado.edu, jeremy.darling@colorado.edu,
charlesdanforth@gmail.com}


\begin{abstract}

Using offset-corrected Gaia-EDR3 parallax measurements and spectrophotometric methods, 
we have determined distances for 69 massive stars in the Carina OB1 association and associated
clusters:  Trumpler~16 (21 stars), Trumpler~14 (20 stars), Trumpler~15 (3 stars), Bochum~11 (5~stars), 
and South Pillars region (20 stars).  Past distance estimates to the Carina Nebula range from 2.2 to 
3.6~kpc, with uncertainties arising from photometry and anomalous dust extinction. The EDR3 
parallax solutions show considerable improvement over DR2, with typical errors 
$\sigma_{\varpi}/\varpi \approx$~3--5\%. The O-type stars in the Great Carina Nebula lie 
at essentially the same distance ($2.35\pm0.08$~kpc), quoting mean and rms variance. The 
clusters have distances of $2.32\pm0.12$~kpc (Tr16), $2.37\pm0.15$~kpc (Tr~14), 
$2.36\pm0.09$ kpc (Tr~15), and $2.33\pm0.12$~kpc (Bochum~11) in good agreement with the 
$\eta$~Car distance of around 2.3 kpc.    O-star proper motions suggest internal 
(3D) velocity dispersions $\sim4$~\kms\ for Tr~14 and Tr~16.  Reliable distances allow estimates of 
cluster sizes, stellar dynamics, luminosities, and fluxes of photoionizing radiation 
incident on photodissociation regions in the region.  We estimate that Tr~14 and Tr~16 have 
half-mass radii $r_{\rm h} = 1.5-1.8$~pc, stellar crossing times 
$t_{\rm cr} = r_{\rm h}/v_{\rm m} \approx 0.7-0.8$~Myr, and two-body relaxation times 
$t_{\rm rh} \approx 40-80$~Myr.  The underlying velocity dispersion for Tr~14, if a bound cluster, 
would be $v_{\rm m} \approx 2.1^{+0.7}_{-0.4}$~\kms\  for $N = 7600^{+5800}_{-2600}$  stars.  
With the higher dispersions of the O-stars, inward drift would occur slowly, on times scales of 
3--6~Myr.  

\end{abstract}

\section{Introduction}

Spanning several square degrees in the Southern sky 
($\ell \approx 287^{\circ}$, $b \approx -0.6^{\circ}$), the Great Carina Nebula (NGC 3372) and
its associated molecular clouds and star clusters are widely investigated as active sites of 
massive star formation in the Galaxy (Walborn \etal\ 2012; Smith \& Brooks 2008). The O-type 
stars in the Carina OB1 association (Walborn 1973a; Humphreys 1978; Smith 2006b) are 
contributors to ionizing radiation, stellar wind mass loss, and other feedback to molecular clouds 
and the interstellar medium.  The Carina region is frequently considered  a laboratory for studying 
the birth and evolution of the most massive stars (Turner \& Moffat 1980; DeGioia-Eastwood \etal\ 
2001; Tapia \etal\ 2003;  Hur \etal\ 2012), including prototypes of updated classifications for the 
O3 stars (Walborn 1971, 1982a) and O2 stars (Walborn \etal\ 2002).  The Carina Nebula and 
surrounding molecular clouds are host to a number of star clusters, including 
Trumpler~14 (Tr~14), Trumpler~15 (Tr~15), Trumpler~16 (Tr~16), Collinder~228 (Col 228), 
Bochum~11 (Bo~11), and the Treasure Chest Nebula.   Illustrative maps of these clusters 
are shown in papers by Alexander \etal\ (2016),  Kiminki \& Smith (2018), Povich \etal\ (2019), 
and Smith \etal\ (2005).  The Carina stellar populations are dominated by Tr 14 and Tr~16
together with Tr~15, a less rich cluster with lower extinction.  Tr 16 is best known for the 
presence of $\eta$ Carinae, a massive, variable star that underwent episodes of brightening 
and eruptive mass loss from 1838--1858 (Humphreys \& Davidson 1984; Davidson \& 
Humphreys  1997, 2012) after beginning its life with initial mass $\sim150~M_{\odot}$.   

Despite its fundamental importance for stellar astronomy, the Carina Nebula continues to
have disparate and controversial values for the distances and ages of its cluster components.  
Early distance estimates include values of 2.5~kpc (Faulkner 1963), 2.7~kpc (Sher 1965), 
and 2.8~kpc (Feinstein 1963).  More recent estimates cover a wider range from 2--4 kpc (distance 
moduli $m_V - M_V = 11.5-13.0$) with some suggestions that Tr~14 and Tr~16  lie at different 
distances (Walborn 1973a; Humphreys 1978; Morrell \etal\ 1988).   More accurate estimates 
of around 2.3~kpc were obtained from geometric methods involving proper motions and 
Doppler velocities of expanding filaments in the $\eta$ Car Homunculus.  Previous estimates for 
$\eta$~Car include $2.2\pm0.2$~kpc (Allen \& Hillier 1993), $2.3\pm0.3$~kpc (Meaburn 1999), 
$2.25\pm0.18$~kpc (Davidson \etal\ 2001), and $2.35\pm0.05$~kpc (Smith 2006a).

Studies of the Carina OB stars are further complicated because lines of sight in this direction 
look down the Carina-Sagittarius spiral arm in a complex of dark clouds and bright nebulosity. 
{\bf Table~1} provides a historical summary of distance estimates for the two main clusters, 
Tr~14 and Tr~16.  Walborn (1973a) placed Tr~14 more distant than Tr~16, but later 
suggested that they had a common distance of 2.8~kpc (Walborn 1982b, 1995).  These 
differences were primarily the result of uncertainties in the reddening corrections necessary 
for accurate spectrophotometric distances\footnote{Considerable evidence shows that the 
interstellar dust extinction is anomalous and variable within the Carina region 
(Feinstein \etal\ 1973; Th\'e \etal\ 1980; Tapia \etal\ 1988; Smith 2002; Carraro \etal\ 2004).  
Ratios of total-to-selective extinction vary from $R_V \approx 3.0-3.5$ in Tr~15 and Tr~16 up
to $R_V = 3.8-4.8$ in Tr~14.  These variations are confirmed by extinction measurements 
from visual and near-infrared photometry (Ma\'iz Apell\'aniz \& Barb\'a 2018) who tabulated 
values of visual extinction $A_{V_J}$ and the corrected magnitudes, 
$V_{J,0} \equiv V_J  - A_{V_J}$ needed to find the distance modulus, DM = $V_{J,0} - M_V$.}.  
Large relative distances between Tr~14 and Tr~16 are unlikely, given the association of
bright O stars and surrounding nebulosity.  Smith (2006b) suggested that Tr~14 might be 
slightly behind Tr~16, because dark silhouetted objects in the foreground of Tr~14 point 
toward $\eta$~Car and Tr~16 (Smith \etal\ 2003).

The goal of  this paper is to obtain more accurate parallax distance estimates for OB stars 
in the Carina Nebula clusters, using values from the Third (early) Gaia Data Release (EDR3) 
of the astrometric mission (Gaia Collaboration 2020).   This project extends our previous study  
(Shull \& Danforth 2019) of parallax and photometric distances for 139 OB stars.  That project 
used Gaia-DR2 parallaxes and benefited from updated spectral types (SpTs) and 
visual/near-IR photometry provided by the Galactic O-Star Spectroscopic Survey (GOSS)
in papers by Sota \etal\ (2014; S14) and Ma\'iz Apell\'aniz \& Barb\'a (2018; MAB18).   With 
the longer time interval of observations, the EDR3 parallax solutions show considerable 
improvement over DR2, as gauged by the parallax-to-error ratio ($\varpi / \sigma_{\varpi})$ 
tabulated in the Gaia archive\footnote{\url {https://gea.esac.esa.int/archive}}.  
In Section 2 we describe our methods for deriving parallax distances, using EDR3 data
together with parallax offsets (Lindegren \etal\ 2020a,b).  We compare these distances
to spectrophotometric distances derived from GOSS data, when available, with occasional 
corrections for binary or multiple systems.   We present our distance measurements for 
69 OB-type stars in the Carina OB1 association and its clusters:  
Tr~16 (21 stars), Tr~14 (20 stars), Tr~15 (3 stars), Bochum~11 (5 stars), and the South Pillars 
region (20 stars) around Collinder~228 and Treasure Chest.  Our primary result is that, 
within errors of parallax measurement, the O-type stars in Car OB1 are at essentially the 
same distance, 2.3--2.5~kpc, in agreement with the $\eta$~Car distance of 
$2.35\pm0.05$~kpc (Smith 2006a).  Section~3 summarizes our results and their
implications for cluster dynamics, stellar absolute magnitudes, and local fluxes of  
photoionizing radiation incident on molecular gas in the Carina Nebula.


\section{Methods and Measurements}

This paper derives both parallax distances and spectrophotometric distances to OB-type
stars in the Carina OB1 association.  We follow the procedures of our recent OB-star 
distance survey (Shull \& Danforth 2019), using updated parallaxes from Gaia EDR3 
and spectrophotometric distances based on photometry and updated SpTs from GOSS.
Our sample contains 69 OB stars, with 61 from GOSS.  Eight non-GOSS stars were added 
to increase the samples for Tr~14, Tr~15, and Treasure Chest.  Past analyses of 
Gaia-DR2 data (Lindegren \etal\ 2018;  Arenou \etal\ 2018;  Brown \etal\ 2018) found
systematic fluctuations in parallaxes relative to the reference frame of distant quasars.  
These offsets appear to be greater toward bright stars (Gaia $G  < 12$) and they 
depend on stellar color and location on the sky.  In our previous distance survey, we 
applied a uniform offset of 0.03~mas to DR2 parallaxes,
the recommended DR2 mean (Lindegren \etal\ 2018).  With the new EDR3 data, we 
computed parallax offsets, $\varpi_{\rm Z}$ (in mas) from the algorithm produced 
by the Gaia team (Lindegren \etal\ 2020b).  Most offsets were based on Gaia 
5-parameter solutions, with only three stars having 6-parameter solutions 
(CPD $-59^{\circ}$~2591, CPD $-59^{\circ}$~2626, HD~93161B).  Although the formal  
Gaia offsets are negative, we follow a convention of quoting positive values in the 
text and tables.  These offsets are added to the EDR3 parallax ($\varpi$) to find the 
corrected distance, $D_{\rm EDR3} = [\varpi + \varpi_{\rm Z}]^{-1}$.  Most stars in 
Tr~14, Tr~15, and Tr~16 had (EDR3) offsets of 0.011--0.017 mas, with only two as large 
as 0.025--0.026~mas.   The five stars in Bochum~11 had offsets of 0.012-0.013~mas, 
and most of the 20 stars in the South Pillars region had offsets of 0.011--0.013 mas.  
One star near the Treasure Chest (CPD $-59^{\circ}$~2661) had a larger offset of
0.026~mas.  

{\bf Table~2} lists EDR3 parallaxes and proper motions with errors from the Gaia
archive, together with the parallax offsets found from the Gaia algorithm.  The stars
are arranged into five groups:  Tr~16, Tr~14, Tr~15, Bochum~11, and Other Stars 
(South Pillars region near Coll~228 and Treasure Chest Nebula).  The formal 
Gaia-EDR3 errors on their parallax solutions listed in Table 2 are typically 3\% to 5\%, 
corresponding to $\varpi / \sigma_{\varpi} = 20-30$.  There may also be systematic 
uncertainties in the parallax zero-point offset (Zinn 2021) and problems with parallax
solutions in close binaries.  Because the relative errors are small, we employed no
special procedures to deal with distance bias from parallax inversion.  The range of
 distances ($\pm 1\sigma$) extends from 
$D_{\rm min} \equiv  [\varpi + \sigma_{\varpi}]^{-1}$ to
$D_{\rm max} \equiv [\varpi -\sigma_{\varpi}]^{-1}$ around our tabulated distance of 
$D_{\rm EDR3} = \varpi^{-1}$.   
We averaged the distances toward stars in several Carina Nebula clusters.  These mean 
distances and rms variances are shown in boldface at the end of each grouping.    

{\bf Table 2} also lists spectrophotometric distances for 69~OB stars in the Car OB1 
association, derived by comparing the star's extinction-corrected magnitude to an 
absolute magnitude, $M_V$, based on its SpT and luminosity class.   The absolute 
magnitudes were taken from Table~11 of Bowen \etal\ (2008), based primarily on 
calculations by Vacca \etal\ (1996).  As we will show below, this absolute magnitude 
scale may need to be revised,  although the corrections remain poorly determined in 
this small sample.  For most stars (61 out of 69) we relied on GOSS spectral classifications 
(S14) and photometry (MAB18).  From these, we used the extinction-corrected magnitude,
$V_{J,0} = V_J - A_{V_J}$ to find the distance, 
$D_{\rm phot} = (10~{\rm pc}) 10^{(V_{J,0} - M_V)/5}$.  For the eight other (non-GOSS) 
stars (HD~93190, HD~303304, Tr~14-3, Tr~14-4, Tr~14-5, Tr~14-6, Tr~14-27,
CPD $-59^{\circ}$ 2661) we relied on photometry from sources in the literature
(Hur \etal\ 2012; Gagn\'e \etal\ 2011; Alexander \etal\ 2016;  Zacharias \etal\ 2013).   
To derive the spectrophotometric distances, we corrected the $V$ magnitudes for extinction 
by $A_V = R_V E(B-V)$, taking $R_V = 3.1$ for two stars in Tr~15 and $R_V = 4.0$ for 
five stars in Tr~14. 

{\bf Figure~1} shows the individual distances for stars in Tr~14, Tr~15, Tr~16, Bo11, and
the South Pillars region, plotted vs.\ the significance level (S/N $\equiv \varpi / \sigma_{\varpi}$) 
of the Gaia-EDR3 parallax measurements.  Values of S/N range from 6.8 to 34.9, with 
means of 19.9 (Tr~16), 25.6 (Tr~15), 23.8 (Tr~14), and 26.2 (Bo~11).  We omitted six stars 
from the Gaia statistical sample, owing to uncertain measurements or missing data:  
two stars in Tr~16 (CPD $-59^{\circ}$ 2635 and CPD $-59^{\circ}$ 2636);  two stars in 
Tr~14 (HD 93161A and Tr 14-5); and two stars in Coll~228 (HD~93146A and HD~93206A).  
From the $D_{\rm phot}$ sample, we omitted three stars:  HD~93162 (WR binary in Tr~16),
Tr~14-5, and HD~93146B because of uncertain data.  Table~2 lists the mean distances and
rms variances of $2.32\pm0.12$~kpc (19 stars in Tr~16);  $2.37\pm0.15$~kpc (18 stars in 
Tr~14); $2.36\pm0.09$~kpc (3 stars in Tr~15); $2.33\pm0.12$~kpc (5 stars in Bochum~11); 
and $2.45\pm0.18$~kpc (Other Stars group). The two stars near the Treasure Chest are at 
$2.35\pm0.08$~kpc and $2.41\pm0.08$~kpc.  The distance variances for stars in these clusters 
are 4--5\%, and weighting by the individual parallax errors gives similar results.   
A few stars in our sample could be non-members, and others may have uncertain distances 
owing to close-binary influence on the parallax solutions.  If we consider only the nine stars 
with SpTs of O5 and earlier, presumed to be the youngest and least dispersed population, 
we find distances of  
$2.32\pm0.08$~kpc (five stars in Tr~16), $2.36\pm0.06$~kpc (three stars in Tr~14), and 
$2.45\pm0.11$~kpc (one star in Tr~15).  Taken as a single ensemble, these nine massive 
O-type stars have a mean distance of $2.35\pm0.08$~kpc (rms variance).  

The spectrophotometric distances are more uncertain, with a broader dispersion than in 
EDR3 parallax.  The mean values are larger than the EDR3 distances by 10\% (Tr~16), 
14\% (Tr~14), 12\% (Bo~11), and 12\% (South Pillars).  Better agreement is found using 
$M_V$ values from Martin \etal\ (2005).  For main-sequence O-type stars, the values in 
Martins \etal\ (2005) are typically lower by $\Delta M_V \approx 0.22-0.28$ mag
(stars are 11--14\% closer in distance) compared to $M_V$ in Vacca \etal\ (1996) and tabulated 
in Bowen \etal\ (2008).  Other discrepancies may arise from incorrect SpTs or difficulties in 
correcting the observed magnitudes of spectroscopic binaries.  Some previous Carina studies 
used uncertain corrections for anomalous extinction and variations in $R_V$.   These corrections 
should be more reliable for the 61 O-type stars drawn from the GOSS sample, where the visual 
extinction $A_{V_J}$ was derived (MAB18) by modeling the visible/near-IR photometry.  For 
the eight non-GOSS stars, larger uncertainties exist in photometry and SpTs.  These issues 
are discussed further in the Appendix for individual stars.

Formal errors on spectrophotometric distances arise from both photometry and the assumed 
absolute magnitude (and SpT).  The GOSS photometry (MAB18) provided values of visual 
extinction $A_{V_J}$ and extinction-corrected visual magnitude 
$V_{J,0} \equiv V_{J} - A_{V_J}$.  The photometric distance is
\begin{equation}
    D_{\rm phot} \equiv (10~{\rm pc}) 10^{(V_{J,0} - M_V) / 5} =
             (10~{\rm pc})  \exp [(\ln 10/5) (V_{J,0} - M_V) ]   \; .
\end{equation}
Taking partial derivatives of $D_{\rm phot}$ with respect to the two independent 
variables, $V_{J,0}$ and $M_V$, we find the propagated distance error,
\begin{equation} 
    \frac {\sigma_D } { D} = \left( \frac {\ln 10} {5} \right)       
    \left[ \sigma_{V_{J,0}}^2  + \sigma_{M_V}^2 \right] ^{1/2}  \; .
\end{equation} 
The typical GOSS errors on $V_{J,0}$ are $\sigma_{V_J} = 0.02-0.03$, while those
on $M_V$ could be as large as $\sigma_{M_V} = 0.15$, corresponding to a half-step 
in SpT (e.g., O5 V to O5.5 V or O8 V to O8.5 V).   For these errors, we find statistical
errors on photometric distance of 7\% ($0.16-0.17$~kpc at $D = 2.3-2.4$~kpc).  

{\bf Figure~2} shows the relation between parallax distance and photometric distances,
$D_{\rm EDR3}$ vs.\ $ D_{\rm phot}$. One immediately sees wide variations in their 
ratio, emphasizing the unreliability of photometric distances in the Carina Nebula.  As 
noted in previous studies, the Carina region exhibits anomalous extinction, with a range of 
values of $E(B-V)$ and $R_V$.  The SpTs and absolute magnitude scales may also need 
recalibration for the massive stars.   {\bf Table 3} compares spectrophotometric distances 
calculated with the two scales.  For many stars, the Martins \etal\ (2005) values provide 
better agreement with EDR3 distances.  The remaining anomalies could be reconciled 
with studies of more stars in other OB associations.

\section{Implications of Distances}

The controversy over the disparate distances ($d$) to Tr~14, Tr~15, and Tr~16 appears to be 
settled.  Within the errors of parallax measurement, the clusters are essentially at the same 
distance, $2.35\pm0.08$~kpc, in good agreement with the previous distance estimates for 
$\eta$~Car:  $2.3\pm0.2$~kpc  (Allen \& Hiller 1993), $2.25\pm0.18$~kpc (Davidson \etal\ 
2001), and $2.35\pm0.05$ kpc (Smith 2006a).  These Gaia-EDR3 distances agree with 
several early recommended values (Th\'e \etal\ 1980; Tapia \etal\ 1988; Walborn 1995), but 
they differ with subsequent estimates of 2.7--3.6~kpc.  
The variations are primarily the result of uncertain and variable dust extinction.  
An accurate distance to the Carina Nebula and its clusters is important for a variety of 
astrophysical calculations for cluster dynamics, stellar properties, and energy feedback to 
the surrounding gas.  For example, cluster radii scale as $R_{\rm c} \propto d$, and 
stellar densities as $n_{\rm c} \propto d^{-3}$.  Cluster masses scale as $M_{\rm c} \propto d^2$, 
when inferred from K-band luminosity functions (Ascenso \etal\ 2007).  As we show, the cluster 
half-mass relaxation time is quite sensitive to distance,  $t_{\rm rh} \propto d^{5/2}$.   Thus, a few
past estimates of the properties of Tr~14, Tr~16, and the Carina OB1 association may need 
evaluation.

Smith (2006b) compiled the cumulative energy input of ionizing photon luminosities, stellar 
mass-loss rates, and mechanical energy production for Carina.  At an assumed distance of 
2.3~kpc for 65 O-type stars and three Wolf-Rayet stars, he found production rates of 
$Q_{\rm H} \approx 10^{51}$~s$^{-1}$ of ionizing (Lyman continuum) photons and 
$L_{\rm w} \approx 3 \times10^{38}$ erg~s$^{-1}$ of stellar-wind mechanical energy.  Because 
his adopted distance is similar to our value, these estimates should be reasonably accurate, 
subject to recent updates of effective temperatures and Lyman-continuum production rates 
from non-LTE, line-blanketed model atmospheres (Pauldrach \etal\ 2001; Puls \etal\ 2005;
Lanz \& Hubeny 2007) and evolutionary tracks for stars with winds and rotation 
(Ekstr\"om \etal\ 2012; Georgy \etal\ 2013).  These effects, and those of metallicity or initial 
mass function, can alter $Q_{\rm H}$ by factors of three for stellar populations 
(Topping \& Shull 2015).  Binary evolution can also affect the time history of $Q_{\rm H}(t)$ 
owing to mass transfer, stellar mergers, and atmosphere tidal stripping (Eldridge \etal\ 2008; 
de Mink \etal\ 2013, 2014).

At a distance of 2.35~kpc and angular scale of 1 arcmin = 0.684 pc, the Carina Nebula 
has a spatial extent of more than 50~pc across the sky.  From the cluster centers and 
circular extents in Tapia \etal\ (2013) we find radii of $4.40$~arcmin (3.0~pc) for Tr~14 
and $5.33$~arcmin (3.65~pc) for Tr~16.  The Tr~14 and Tr~16 centers are separated by 
$19.7$~arcmin (13.5~pc),  and Tr~15 and Tr~14 by $16.5$~arcmin (11.3~pc).   These
sizes allow analysis of the dynamical evolution of the star clusters. Trumpler~14 is  one 
of the youngest structures, with an estimated, but uncertain age of 
$t_{\rm cl} \approx 0.5-1.0$~Myr  (Walborn 1995; Tapia \etal\ 2003).  From a near-infrared 
survey and Kroupa (2001) initial mass function, Ascenso \etal\ (2007) estimated a cluster 
mass $M_{\rm c} \approx (9000~M_{\odot})d_{2800}^2$ (they assumed a distance of 
2800~pc) with a substantial portion coming from pre-main-sequence (PMS) stars.  Using 
near-infrared multi-conjugate adaptive optics on the VLT, Sana \etal\ (2010) modeled the 
spatial distribution of stars within Tr~14 to derive  
$M_{\rm c} = (4.3^{+3.3}_{-1.5} \times 10^3\,M_{\odot})d_{2500}^2$ (they assumed a distance
of 2500~pc).  Rescaling their numbers to $d = (2350~{\rm pc}) d_{2350}$ and assuming a 
mean stellar mass  $m_{\rm f} \approx 0.5~M_{\odot}$, we find that Tr~14 has mass 
$M_{\rm c} = (3.8^{+2.9}_{-1.3} \times 10^3\,M_{\odot})d_{2350}^2$ 
and $N = (7600^{+5800}_{-2600})d_{2350}^2$ stars within a half-mass radius of
$r_{\rm h} = (1.5~{\rm pc})d_{2350}$.  We adopt an (rms) velocity dispersion for the underlying
cluster stars of  $v_{\rm m} \approx (2.1^{+0.7}_{-0.4}$~\kms)$d_{2350}^{1/2}$, assuming the
relation for a bound cluster, $v_{\rm m}^2 \approx 0.4 (GM_{\rm c}/r_{\rm h}$), from Spitzer (1987).  
The average stellar density (at $r \leq r_{\rm h}$) is 
$n_{\rm h} = (3/8 \pi)(N/r_{\rm h}^3) \approx (270^{+200}_{-95}~{\rm pc}^{-3})d_{2350}^{-1}$.   
From these parameters, we estimate two relevant dynamical times for Tr~14,
\begin{eqnarray}
   t_{\rm cr} & = & r_{\rm h} / v_{\rm m} \approx (0.70^{+0.16}_{-0.18}~{\rm Myr}) \, d_{2350}^{1/2}    \\
   t_{\rm rh} &=& \frac {v_{\rm m}^3}{1.22 n_{\rm h} (4 \pi G^2 m_{\rm f}^2 \ln \Lambda)} 
                = (58^{+19}_{-11}~{\rm Myr}) \, d_{2350}^{5/2}   \; ,
\end{eqnarray}  
for the cluster crossing time and the half-mass relaxation time (Spitzer 1987).  We have
taken the long-range logarithm, $\ln \Lambda \approx \ln (0.4 N) \approx 8$.  The scaling 
with distance follows from $r_{\rm h} \propto d$, $M_{\rm c}  \propto d^2$, 
$n_{\rm h} \propto  N/r_{\rm h}^3 \propto d^{-1}$, and 
$v_{\rm m} \propto (GM_{\rm c}/r_{\rm h})^{1/2} \propto d^{1/2}$.  
The two-body relaxation time of the cluster has the greatest sensitivity to distance, with 
$t_{\rm rh} \propto (v_{\rm m}^3 /n_{\rm h}) \propto d^{5/2}$.  

Given the young age of Tr~14, most stars have probably undergone only one or two orbits within 
the cluster.  The cluster may have expanded as gas
was blown out, and a few stars may have moved to outer portions of the cluster where they 
could be stripped tidally.  Young OB associations are believed to be slowly expanding, with 
5--30\% of their stars ejected as high-velocity stars (Gies 1987; Hoogerwerf \etal\ 2000). 
Since $t_{\rm rh}$ is over 50 times the cluster age, most stars in Tr~14 are unlikely to have 
experienced significant overall relaxation effects (stellar dynamical escape, core collapse).  
Those effects occur on timescales of  $(2-3) t_{\rm rh}$, as found in multi-mass component
models of cluster evolution (Spitzer \& Shull 1975).  The most massive stars in the cluster 
could drift inward through dynamical friction (Binney \& Tremaine 2008) as discussed by 
Sana \etal\ (2010).  Inward drift of an O-type star of mass $M_* = (40~M_{\odot}) M_{40}$ 
would occur on a time scale
\begin{equation}
   t_{\rm df} = \frac {V_*^3} {n_{\rm f}  m_{\rm f} M_* (4 \pi G^2 \ln \Lambda)} \approx
         (6.2~{\rm Myr}) V_4^3 M_{40}^{-1}    \; .
\end{equation} 
Here, we scaled the massive star velocity $V_*$ to the 3D velocity dispersion, 
$(4~{\rm km~s}^{-1}) V_4$,  found from O-star proper motions internal to
Tr~14 and Tr~16 (see below).  Thus, the massive stars ($30-80~M_{\odot}$)
in these clusters should have undergone little mass segregation over their
1--3~Myr lifetimes.  

The Tr~16 cluster is slightly larger than Tr~14 and slightly more dispersed.  
Previous surveys (DeGioia-Eastwood \etal\ 2001;  Wolk \etal\ 2011) suggested that 
its stellar numbers are similar.  From source counts, Wolk \etal\ (2011) found 1232 
X-ray sources, matched to 1187 likely near-IR members, and estimated a PMS 
population of  $>6500$ class~II-III objects.  We conservatively adopt a total 
$N \approx 8000$.   With $r_{\rm h}  \approx 1.8$~pc, we estimate 
$n_{\rm f} \approx 160$~pc$^{-3}$, $M_{\rm c} \approx 4000~M_{\odot}$ 
(with $m_{\rm f} = 0.5~M_{\odot}$), $v_{\rm m} \approx 2.0$~\kms, 
$t_{\rm cr} \approx 0.8$~Myr, and $t_{\rm rh} \approx 80$~Myr.  
The stellar velocity dispersions of Tr~14 and Tr~16 are still uncertain, with 
estimates from both radial velocity (RV) surveys and EDR3 proper motions. 
Young clusters will expand as they blow away their nascent cloud material.
The velocity dispersion of the O-type stars may exceed our 2.1~\kms\ estimate for the 
underlying stellar population (including PMS stars) assuming a gravitationally bound 
system. From a survey of the well-constrained stars in Tr~14, Kiminki \& Smith (2018) 
found a weighted-mean heliocentric RV =  $2.3 \pm 7.4$~\kms, comparable to the
value, $2.8 \pm 4.9$~\kms, found by Penny \etal\ (1993) for six O stars.  (The total
velocity dispersion would be $\sqrt{3}$ times the 1D value.)  They regard their 
7.4~\kms\ dispersion as an upper limit, since it is larger than dispersions of typical, 
less massive OB associations.

{\bf Table 4} lists the proper motions (PM) for individual stars in Carina OB1, for which
the Gaia-reported motions (in the Sun's frame) are dominated by the RA component.  
To visualize the internal motions, we transformed PMs to the Carina frame 
by subtracting the mean values for each cluster.  We list values of PM in both 
the Sun's frame and the cluster rest frame.  At 2.35~kpc distance, a PM of 
1~mas~yr$^{-1}$ corresponds to transverse velocity 11.14~\kms\ on the sky.  
In Table 2, we found a mean total PM  for Tr~14 of of $6.944\pm0.221$~mas~yr$^{-1}$ 
(Sun's frame) quoting the rms variance among 20 stars.  The PM directional components 
have mean values 
${\rm PM}_{\rm RA} = -6.944\pm0.261$~mas~yr$^{-1}$ in right ascension (RA) and 
${\rm PM}_{\rm Dec} = 2.185\pm0.366$~mas~yr$^{-1}$ in declination (Dec), similar 
to the proper motions found by Kuhn \etal\ (2019) from a larger number of lower-mass
stars.  In the Gaia archive tables, 
${\rm PM}^2 = ({\rm PM}_{\rm RA})^2 + ({\rm PM}_{\rm Dec})^2$. 
Table~4 shows the components in RA and Dec, yielding the total PM amplitude.  
Its propagated uncertainty was found from errors in the individual RA and Dec 
measurements (typically $\pm 0.014-0.027$~mas~yr$^{-1}$) and the uncertainty on the 
cluster mean PMs.   We interpret  the internal PM amplitude as the (2D) internal velocity
dispersion for each cluster.  For an isotropic velocity distribution, the 3D dispersion is 
higher by a factor $\sqrt{3/2}$.   We derive PM errors for each star from the component 
errors, $\sigma_{{\rm PM}_{\rm RA}}$ and $\sigma_{{\rm PM}_{\rm Dec}}$, 
\begin{equation}
   \sigma_{\rm PM}^2 =  \left( \frac {{\rm PM}_{\rm RA}}{{\rm PM}} \right)^2 
         \sigma_{{\rm PM}_{\rm RA}}^2  + \left( \frac {{\rm PM}_{\rm Dec}}{{\rm PM}} \right)^2 
         \sigma_{{\rm PM}_{\rm Dec}}^2   \; .
\end{equation}
An interesting ancillary result from the mean values in Table 4 is that Tr~16 appears to 
be approaching Tr~14 by 0.351~mas~yr$^{-1}$ in RA and 0.427~mas~yr$^{-1}$ in 
declination (6~\kms\ total motion). These values are larger than statistical errors 
on the mean component PMs.  

The 20 OB stars in Tr~14 have a mean 2D internal dispersion of 
$\sigma_{\rm PM} =  0.359\pm0.058$~mas~yr$^{-1}$.  Two stars (HD~93160 and 
ALS~15207) have high PMs (0.848~mas~yr$^{-1}$), more than twice the cluster mean 
and thus candidates for escape.  If we omit those two stars from the sample, the mean 
dispersion drops to $\sigma_{\rm PM} =  0.304\pm0.049$~mas~yr$^{-1}$, 
corresponding to velocity dispersions of 3.4~\kms\ (2D) and 4.1~\kms\ (3D).  The 20 stars 
in Tr~16 have a mean 2D dispersion $\sigma_{\rm PM} =  0.311\pm0.044$~mas~yr$^{-1}$, 
corresponding to velocity dispersions of 3.5~\kms\ (2D) and 4.2~\kms\ (3D).  Two stars 
(HD~93205 and HD~303316) have high PMs (0.765 and 0.749~mas~yr$^{-1}$) and
are escape candidates.  Omitting those two stars from the sample reduces the
dispersion to $\sigma_{\rm PM} =  0.262\pm0.013$~mas~yr$^{-1}$.  The group of
five O-type stars in Bochum~11 has the smallest mean dispersion,
$\sigma_{\rm PM} =  0.093\pm0.023$~mas~yr$^{-1}$, corresponding to a 3D velocity 
dispersion of 1.3~\kms.  The dispersions for the massive OB-type stars suggest that Tr~14
and Tr~16 may not be in virial equilibrium and in the process of dispersing (Kuhn \etal\ 2019).
Multi-body interactions between binaries and cluster stars could  transfer energy to these 
stars and produce expansion of the cluster.

Although Tr~14 has a core-halo structure (Ascenso \etal\ 2007), no clear evidence has
been found for mass segregation (Sana \etal\ 2010).  Proper motions shows that the 
massive stars have velocities above the assumed value ($v_{\rm m} \approx 2$~\kms) from 
the virial equilibrium relation, $v_{\rm m} ^2 \approx 0.4 (GM_{\rm c} /r_h)$, adopted by
Spitzer (1987).   The OB stars may already be expanding from the cluster, with velocities 
of 3--4~\kms\ relative to the low-mass PMS stars that dominate two-body relaxation.  
Inward drift from dynamical friction likely occurs slowly, over 3--6~Myr, and it might be 
counteracted by expansionary processes such as gravitational tides, multi-body interactions 
of stars, and supernova explosions in binary systems.   The latter two processes could produce 
runaway stars (Blauuw 1961; Gies 1987; Hoogerwerf \etal\ 2000; Bally \& Zinnecker 2005).

\section{Summary}

An encouraging result of our study is that EDR3 parallaxes show considerable 
improvement over DR2, yielding reliable distances in most cases.   Table 2 lists the error 
range ($D_{\rm min}, D_{\rm max}$) about the mean parallax distances $D_{\rm EDR3}$.  
With parallax errors, $\sigma_{\varpi} \approx 0.013-0.020$~mas, most OB stars in our 
Car~OB1 survey have fractional errors, $\sigma_{\varpi}/\varpi \approx$ 3.5\% to 5.0\%, 
smaller than the uncertainties in the spectrophotometric distances.  These parallax errors 
correspond to radial distance uncertainties of 40--100~pc, comparable to the extent of the 
Carina Nebula on the sky.  The main conclusions and implications of our study are as follows:
 \begin{enumerate}
   
 \item   For the youngest stellar populations (O5 and earlier), we find distances of 
 $2.32\pm0.08$~kpc (five stars in Tr~16); $2.36\pm0.06$~kpc (three stars in Tr~14); and 
 $2.45\pm0.11$~kpc (one star in Tr~15) quoting rms variances. Taken as a single ensemble, 
 these nine stars have a mean distance of $2.35\pm0.08$~kpc (rms). 

\item The uniformity in parallax distances to Tr~14, Tr~15, Tr~16, Bo~11, and Treasure Chest
  is consistent with observations  of anomalous and variable dust extinction within the Carina 
  Nebula.  Comparisons of EDR3 distances to new spectrophotometric parallaxes confirm the 
  high values of $R_V \approx 4$ for Tr~14 compared to Tr~16 and Tr~15.   The current survey 
  is consistent with the prescient statement (Walborn 1995) in his review of the Carina Nebula:
 ``If $R$(Tr 14) = $R$(Tr 16) + 1, there would be no need for either a distance or age 
 difference between the two clusters".  

 \item An established distance ($d \approx 2.35$~kpc) to the Carina Nebula provides 
 confidence in the previous calibration of stellar luminosities, FUV fluxes, and mass-loss 
 rates (Smith 2006b) based on $d = 2.3$~kpc.  With new model atmospheres, these may 
 shift together with new stellar scales of $T_{\rm eff}$,  $M_V$, and ionizing photon production.  
 These parameters determine the photoevaporation rates of protoplanetary disks and 
 properties of gas pillars and photodissociation regions (Smith \& Brooks 2008).  

\item  With $d = 2.35$~kpc, the cluster two-body relaxation time, $t_{\rm rh} \propto d^{5/2}$, is
reduced significantly from previous values at 2.5--2.8~kpc distances.  The star clusters appear 
dynamically young.  With radii of 3--4~pc, Tr~14 and Tr~16 have stellar crossing times 
$0.7-0.8$~Myr and relaxation times $t_{\rm rh} \approx 40-80$~Myr.   Proper motions of the 
O-type stars in Tr~14 and Tr~16 suggest velocity dispersions of 3-4~\kms.  Massive stars
could drift inward by dynamical friction on 6~Myr times scales.  Thus, mass segregation and
inward drift are unlikely to be observable for these dynamically young clusters.  

\item We identified stars in Tr~14 (HD~93160 and ALS 15207) and Tr~16 (HD~93205 and  
HD~303316) with proper motions more than twice the cluster dispersion.  These stars could 
be candidates for escape, produced by multi-body stellar interactions.   Although no obvious 
supernovae have been found in the Carina Nebula, runaway stars from dense cluster cores 
would travel 20--100~pc over 1~Myr.  

 \item Most parallax distances in Car~OB1 are more accurate than spectrophotometric estimates 
 found from GOSS photometry and SpTs.   Mean photometric distances to Tr~14, Tr~16, and  Bo~11 
 are 10-14\% larger than Gaia-EDR3 distances.  A set of massive stars at a common distance
 with reliable reddening models (e.g., GOSS) could enable a calibration of absolute magnitudes.
For most main-sequence O-stars, the Martins \etal\ (2008) scale of $M_V$ gives somewhat
better agreement with EDR3 distances.
 
 \end{enumerate}
 
 
\vspace{0.5cm}

\noindent
{\bf Acknowledgements.}
We thank Nathan Smith, Roberta Humphreys, John Bally, for helpful discussions about the 
Carina Nebula and Kris Davidson and Bill Vacca for comments on distance measurements. 
This work has made use of data from the European Space Agency (ESA) mission Gaia 
(\url{https://www.cosmos.esa.int/gaia}), processed by the Gaia Data Processing and Analysis 
Consortium (DPAC, \url{https://www.cosmos.esa.int/web/gaia/dpac/consortium}). Funding for the 
DPAC has been provided by national institutions, in particular the institutions participating in the 
Gaia Multilateral Agreement.

\appendix

\section{Notes on Individual Stars}

\noindent
{\bf CPD $-59^{\circ}$ 2603.}   This is a hierarchical triple system, also known as V572 Car, 
with spectral classification [O7.5~V + B0~V] in Table 7 of S14.  An alternate classification 
of  [O7~V + O9.5~V + B0.2~IV] comes from high-resolution spectra (Rauw \etal\ 2001).
From GOSS photometry (MAB18) of the full system, we use $V_{J,0} = 7.17$ and adopt 
$M_V = -4.75$ (for O7~V) to find $D_{\rm phot} = 2.42$~kpc.  This distance is slightly less
than the Gaia-EDR3 value of 2.62~kpc (range 2.48-2.78 kpc).  

\noindent
{\bf CPD $-59^{\circ}$ 2624.}   Also known as Tr~16-9, this star was incorrectly listed as
CPD-$59^{\circ}2634$ in the GOSS papers (S14 and MAB18).  The SpT is O9.7~IV (S14)
for which $M_V = -4.56$ for O9.7~IV (Bowen \etal\ 2008).  We adopt an extinction 
corrected $V_{J,0} = 7.603\pm0.016$ (MAB18) to find $D_{\rm phot} = 2.71$~kpc.  
This photometric distance would decrease to 2.24~kpc if we used the O9.5~V SpT of 
Nelan \etal\ (2004) and $M_V = -4.15$.   This shorter distance would be in better 
agreement with the Gaia value, $D_{\rm EDR3} = 2.26$~kpc (range 1.98--2.64)~kpc.  
 
 \noindent
{\bf CPD $-59^{\circ}$ 2635.}  This is a spectroscopic binary, classified by GOSS (S14) 
as [O8~V + O9.5~V] with an extinction-corrected magnitude (MAB18) of 
$V_{J,0} = 6.844\pm0.019$ with $A_{V_J} = 2.437\pm0.025$. 
Adopting $M_V = -4.60$ for O8 V (Bowen \etal\ 2008) we find 
$D_{\rm phot} = 2.50$~kpc, after correcting the distance to the two stars with 
a multiplicative factor $[1 + (L_2/L_1)]^{1/2} = 1.29$ for $L_2/L_1 = 0.661$.   
 
\noindent
{\bf CPD $-59^{\circ}$ 2636AB.}  Also known as Tr 16-110, this star had no parallax 
solution in either DR2 or EDR3 of the Gaia archive.  It appears in the GOS survey (S14) 
as a spectroscopic binary [O8~V + O8~V] with an extinction-corrected magnitude 
(MAB18) of $V_{J,0} = 6.493\pm0.019$ with $A_{V_J} = 2.726\pm0.021$. 
With $(B-V) = 0.34$ and $(B-V)_0 = -0.31$, we find $E(B-V) = 0.65$, consistent with 
$R_V \approx 4$.  Adopting $M_V = -4.60$ for O8 V (Bowen \etal\ 2008) we find 
$D_{\rm phot} = 2.34$~kpc, after correcting the distance to the two O8~V stars with 
a multiplicative factor $[1 + (L_2/L_1)]^{1/2} = \sqrt 2$.  The lack of a parallax solution
may arise because the system is actually quadruple  (S14).  

\noindent
{\bf HD 93129AB.}  GOSS (S14) lists both A and B stars as HD~93129AaAb (O2~If*) 
and HD~93129B [O3.5~V((f))].   GOSS photometry (MAB18) 
lists an  extinction-corrected magnitude $V_{J,0} = 4.825\pm0.029$ with 
$A_{V_J} = 2.199\pm0.025$ for the combined system (AaAbB).  
The AaAb binary was classified by Gruner \etal\ (2019) as [O2 If* + O3 III(f*)]
with $V_{\rm Aa} = 7.65$, $V_{\rm Ab} = 8.55$, and $(B-V) = 0.25$.  Using
$(B-V)_0 = -0.32$, we adopt $E(B-V) = 0.57$ for the system (Aa, Ab, and B) and 
a combined magnitude $V_{\rm AaAb} = 7.26$.  With a combined $M_V = -6.49$ for 
Aa+Ab, we find $D_{\rm phot} = 2.49$~kpc, a distance that we assign to both A and B.  

\noindent
{\bf HD 93146AB.}  This is a multiple system located in the Collinder~228 region.  
GOSS (S14) tabulates two separate stars, an SB1 spectroscopic binary 
HD~93146A (O7~Vfz) and HD~93146B (O9.7~IV).  GOSS photometry (MAB18) 
lists extinction-corrected magnitudes,
$V_{J,0}(A) = 6.905\pm0.027$ with $A_{V_J} = 1.535\pm0.029$ and
$V_{J,0}(B) = 8.602\pm0.027$ with $A_{V_J} = 1.324\pm0.046$.  Adopting
$M_V(A) = -4.90$ (O7~V) and $M_B(B) = -4.56$ (O9.7~IV), we derive photometric 
distances of 2.30~kpc (A) and 4.29~kpc (B), considerably different from the 
offset-corrected Gaia-EDR3 parallax distances, 2.93~kpc (A) and 2.47~kpc (B).  
These discrepancies are puzzling, since A and B are separated by just $6.5''$ 
(Figure 12a of S14).  The GOSS magnitude, $V_{J} = V_{J,0}+A_{V_J} = 9.926\pm0.053$ 
is slightly fainter than $V = 9.864$ (Gagn\'e \etal\ 2011) whose $(B-V) \approx 0.00$
indicates $E(B-V) \approx 0.31$.  This color excess is consistent with the GOSS value,
$A_{V_J}/R_V \approx 0.33$. The SpT of HD~93146B may be wrong, leading to an 
incorrect $M_V$, although it would need to be early B-type.  Alternately, HD~93146B 
could be a background star.  

\noindent
{\bf HD 93161AB.}   
GOSS (S14) tabulates A and B stars separately: an SB2 binary HD~93161A
[O7.5~V + O9~V] and HD~93161B (O6.5~IV).  However, GOSS photometry (MAB18)
lists extinction-corrected magnitudes for the full system (AB) with 
$V_{J,0} = 5.776\pm0.017$ with $A_{V_J} = 2.038\pm0.026$. To obtain an absolute 
magnitude for the AB system, we combine $M_V = -4.75$ (O7.5~V) with 
$M_V = -4.30$ (O9~V) to obtain $M_V(A) = -5.30$, and then with $M_V(B)= -5.40$ 
(O6.5~IV) to find $M_V(AB) = -6.10$.  This yields a system photometric distance of 
$D_{\rm phot} = 2.37$~kpc.   Naz\'e \etal\ (2015) provided individual photometry
and modeling, with $R_V = 3.8$, $V_A = 8.56\pm0.02$ and $V_B = 8.60\pm0.02$.
Color indices $(B-V) = 0.20$ (A) and 0.23 (B) correspond to $E(B-V)$ of 0.52 (A) and 
0.55 (B). The combined magnitude is $V_{AB} = 7.83\pm0.02$, similar to the GOSS 
value of $7.81\pm0.03$ (MAB18).  With $M_V(A) = -5.21$ for their [O8~V + O9~V] 
classification and $M_V(B) = -5.40$ for O6.5~IV, we obtain photometric distances of 
2.28~kpc (A) and 2.41~kpc (B), adopting their value of $R_V = 3.80$.  The Gaia-EDR3 
archive lists different parallaxes for component A ($0.3399\pm0.0281$~mas) and 
component B ($0.3859\pm0.0282$~mas).  
After offset corrections, those correspond to EDR3 distances of 2.84~kpc (A) and 
2.48~kpc (B).  This suggests potential problems with the parallax solution
from system multiplicity.  

\noindent
{\bf HD 93190.}  This star appeared in GOSS with an uncertain O9.7:V: classification
(S14), but no GOSS photometry was found (MAB18).   We adopt $B = 9.10\pm0.02$ and 
$V = 8.84\pm0.04$ (Zacharias \etal\ 2013), from which we estimate $E(B-V) = 0.56$ from 
$(B-V) = 0.26$ and $(B-V)_0 = -0.30$.  With $M_V = -4.00$ for this SpT, we find an uncertain  
spectrophotometric distance of 1.66~kpc, considerably lower than the Gaia-EDR3 
value of 2.35~kpc (range 2.25--2.45 kpc).  The uncertain SpT makes this an unreliable distance.

\noindent
{\bf HD 93206A.}  The variable QZ~Car  is the brightest object in the Collinder~228 
cluster and ID\#46 in our distance survey (Shull \& Danforth 2019).  This is a double-line 
(SB1+SB1) binary (Parkin \etal\ 2011), consisting of system~A (O9.7~Ib + b2 v) and 
system B (O8~III  + o9 v).  GOSS gives a combined classification of O9.7~Ib (S14) with 
$V_{J,0} = 4.206\pm0.018$ and $A_{V,J} = 2.106\pm 0.022$ (MAB18),  
presumably including all four stars in HD 93206AB.    Our photometric distance 
estimate, $D_{\rm phot} = 2.23$~kpc, is based on the Aa component (O9.7~Ib) with 
$V = 6.31$ and $M_V = -6.18$, corrected for the luminosity ratio $L_2/L_1 = 0.535$ 
of the two brightest stars (O8~III and O9.5~Ib), increasing the distance by a factor 
$[1+(L_2/L_1)]^{1/2} = 1.24$.  

\noindent
{\bf HD 93222.}  GOSS (S14) classifies this star as O7~Vfz, and photometry (MAB18) 
has $V_{J,0} = 6.265\pm0.019$ and $A_{V,J} = 1.837\pm0.021$.  The O7~V classification 
with $M_V = -4.90$ would yield an unrealistically small spectrophotometric distance of
$D_{\rm phot} =  1.71$~kpc.  Instead, as recommended
by Shull \& Danforth (2019) for this star (ID \#47), we adopt  O7~IIIf from 
Markova \etal\ (2011) and Walborn (1973b) with $M_V = -5.70$, yielding 
$D_{\rm phot} = 2.47$~kpc.  

\noindent
{\bf HD 303304.}  This star did not appear in GOSS (S14), and no photometry 
was found (MAB18).  The Simbad website lists an uncertain SpT  of O7-8 V and
photometry of $B = 10.08\pm0.20$ and $V = 9.67\pm0.32$ from Alexander \etal\ (2016).
From this, we estimate $E(B-V) \approx 0.73$ from $(B-V) = 0.41$ and $(B-V)_0 = -0.32$.
We adopt $M_V = -4.75$, intermediate between $-4.90$ (O7~V) and $-4.60$ (O8~V), for 
a spectrophotometric distance 2.70~kpc, somewhat higher than the Gaia-EDR3 value 
of 2.28~kpc (range 2.21--2.35 kpc).  Because of the uncertain photometry, reddening, and
SpT, this is an unreliable distance.

\noindent
{\bf HD 303311.}  GOSS photometry (MAB18) listed extinction-corrected 
$V_{J,0} = 7.448\pm0.018$ and $A_{V,J} = 1.523\pm0.024$, and S14 classify it as
O6~V((f))z.  With $M_V = -5.20$ (Bowen \etal\ 2008) we find $D_{\rm phot} = 3.39$~kpc.  
Walborn (1982a) classifies the star as O5~V, while Alexander \etal\ (2016) list it as 
O8~III((f)), with alternate SpTs of O7~V or O6~V((f))z and a possible spectroscopic binary.  
The Gaia-EDR3  parallax distance is 2.24~kpc (range 2.17--2.31~kpc).  The O8~III 
classification ($M_V = -5.50$) would lead to a distance of 3.89~kpc.  Better agreement 
between $D_{\rm phot}$ and $D_{\rm EDR3}$ would be found with a later SpT.
For example, a main-sequence classification of O8~V with $M_V = -4.60$ would give 
$D_{\rm phot} = 2.57$~kpc.  

\noindent
{\bf HD 305518.}  GOSS photometry (MAB18) lists extinction-corrected 
$V_{J,0} = 7.284\pm0.022$ and $A_{V,J} = 2.435\pm0.032$, and S14 classify it as
O9.7 III.  With $M_V = -5.10$ (Bowen \etal\ 2008) we find $D_{\rm phot} = 3.00$~kpc,
which is considerably larger than $D_{\rm EDR3} = 2.00$~kpc (range 1.92--2.08~kpc). 
This star is fairly distant from Tr~16 on the sky and loosely associated with Col~228.  
If the DR3 parallax is reliable, the star could be in the foreground.  It is difficult
to decide, given the disparity between the two distances.  

\noindent
{\bf HD 305532.}  This star is located near the Treasure Chest, with an EDR3 distance 
of $2.41\pm0.08$~kpc.    GOSS photometry (MAB18) lists extinction-corrected 
$V_{J,0} = 7.216\pm0.023$ and $A_{V,J} = 2.946\pm0.040$, and S14 classify it as
O6.5 V((f))z.  With $M_V = -5.05$ (Bowen \etal\ 2008) we find $D_{\rm phot} = 2.84$~kpc,
which is greater than $D_{\rm EDR3} = 2.41$~kpc (range 2.33--2.49~kpc).  The
star may be an unresolved SB1 binary (Levato \etal\ 1990).  However, a standard 
luminosity correction would worsen the discrepancy, increasing the photoelectric distance 
by a factor  $[1 + (L_2/L_1)]^{1/2}$.  This suggests that the star may have a fainter $M_V$
than we adopted for its SpT of O6.5~V in GOSS.  

\noindent
{\bf CPD $-59^{\circ}$ 2661.}  This star is located near the Treasure Chest, with an EDR3
distance of $2.35\pm0.08$~kpc. This star did not appear in GOSS (S14), and no 
photometry was found (MAB18).  Walsh (1984) found a SpT of O9.5~V, and photometry 
of $B = 11.55\pm0.16$ and $V = 11.20\pm0.18$ comes from Zacharias \etal\ (2013).  From 
this, we estimate $E(B-V) \approx 0.66$ from $(B-V) = 0.35$ and $(B-V)_0 = -0.31$, in 
agreement wth $E(B-V) = 0.65\pm0.04$ and $R_V = 4.8$ from hydrogen recombination line 
ratios (Smith \etal\ 2005).  With $M_V = -4.15$ (Bowen \etal\ 2008), we find 
$D_{\rm phot} = 2.79$~kpc, with at least 10\% uncertainty from photometry errors.  This 
suggests that the star may have a fainter $M_V$ than we adopted for its SpT of O9.5~V.  

\noindent
{\bf Tr 14-3.}  This star did not appear in GOSS  (S14), and no photometry was found 
(MAB18).  Nelan \etal\ (2004) list a SpT of B0.5~IV-V, originally classified by 
Morrell \etal\ (1988).  Levato \etal\ (2000) found that Tr~14-3 is a double-lined 
spectroscopic binary, with approximate masses of $20~M_{\odot}$ and $15~M_{\odot}$.  
Hur \etal\ (2012) quoted  $B = 11.00$ and $V = 10.73$.  With $(B-V) = 0.27$ and 
$(B-V)_0 = -0.28$, we find $E(B-V) = 0.55$.  We adopt $M_V = -3.88$, intermediate 
between $-3.55$ (B0.5~V) and $-4.20$ (B0.5~IV).  If $R_V = 4.0$, the spectrophotometric 
distance would be 3.03~kpc.  Instead, we adopt a SpT of B0.5~V, under the assumption
that the B stars in Tr14 should still be on the main sequence.  We quote an uncertain 
distance $D_{\rm phot} = 2.61$~kpc, close to the Gaia-EDR3 distance of 2.42~kpc
(range 2.35--2.49~kpc).  

\noindent
{\bf Tr 14-4.}  This star did not appear in GOSS  (S14), and no photometry was found 
(MAB18).   Gagn\'e \etal\ (2011) classified this star as B0~V, and Hur \etal\ (2012) quoted 
$B = 11.32$ and $V = 11.04$.  With $(B-V) = 0.28$ and $(B-V)_0 = -0.30$, 
we have $E(B-V) = 0.58$.  With $M_V =  -4.30$ and $R_V = 4.0$, we find an uncertain
$D_{\rm phot} = 3.27$~kpc, considerably higher than the Gaia-EDR3 distance of 
2.62~kpc (range 2.54--2.71~kpc).  An incorrect SpT, higher extinction, or $R_V \geq 4.5$ 
could all account for the difference.

\noindent
{\bf Tr 14-5.}  This star did not appear in GOSS (S14), and no photometry was found 
(MAB18).  Gagn\'e \etal\ (2011) classified this star as O9~V, and Hur \etal\ (2012) quoted 
$B = 11.63$ and $V = 11.30$.  With $(B-V) = 0.33$ and $(B-V)_0 = -0.31$, we find 
$E(B-V) = 0.64$.  With $M_V =  -4.30$ and $R_V = 4.0$, we find an uncertain
$D_{\rm phot} = 4.06$~kpc, considerably higher than the Gaia-EDR3 distance of 
2.83~kpc (range 2.71--2.96~kpc).   An incorrect SpT, higher extinction, or $R_V \geq 4.5$ 
could all account for the difference.

\noindent
{\bf Tr 14-6.} This star did not appear in GOSS (S14) and no photometry was found 
(MAB18).  Gagn\'e \etal\ (2011) classified this star as B1~V, and Hur \etal\ (2012) quoted 
$B = 11.33$ and $V = 11.12$.  With $(B-V) = 0.21$ and $(B-V)_0 = -0.26$, we find 
$E(B-V) = 0.47$.  With $M_V =  -3.04$ and $R_V = 4.0$, we find an uncertain 
$D_{\rm phot} = 2.86$~kpc, higher than the Gaia-EDR3 distance of 
2.26~kpc (range 2.18--2.34~kpc).  

\noindent
{\bf Tr 14-27.}  This star did not appear in GOSS (S14), and no photometry was found 
(MAB18).   Alexander \etal\ (2011) classified this star as B1.5~V, and Hur \etal\ (2012) 
quoted $B = 11.54$ and $V = 11.22$.  With $(B-V) = 0.32$ and $(B-V)_0 = -0.25$, 
we estimate $E(B-V) = 0.57$.  With $M_V =  -4.30$ and $R_V = 4.0$, find
$D_{\rm phot} = 2.09$~kpc, less than the Gaia-EDR3 distance of 2.21~kpc  
(range 2.15--2.28~kpc).  

\noindent
{\bf ALS~15210.}  This star was classified as O3.5~If* in(S14), but no GOSS photometry 
was provided (MAB18).  Hur \etal\ (2012) quoted $B = 11.507$ and $V = 10.709\pm0.013$.  
With $(B-V) = 0.80$ and $(B-V)_0 = -0.26$, we estimate $E(B-V) \approx 1.12$.  The 
correction factor $R_V = A_V/E(B-V)$ is likely large but uncertain.  We  adopt $R_V = 4.5$ 
and $A_V \approx 5.0$~mag, similar to the GOSS value (MAB18) for the nearby star 
HD~93162.  With $M_V =  -6.30$ for O6~Ib (Bowen \etal\ 2008) we find
$D_{\rm phot} = 2.48$~kpc, similar to the Gaia-EDR3 distance of 2.40~kpc  
(range 2.32--2.48~kpc).  This photometric distance comes with considerable
uncertainty owing to the reddening correction.


\newpage


\begin{figure}[!ht]
\includegraphics[angle=0,scale=1.05]{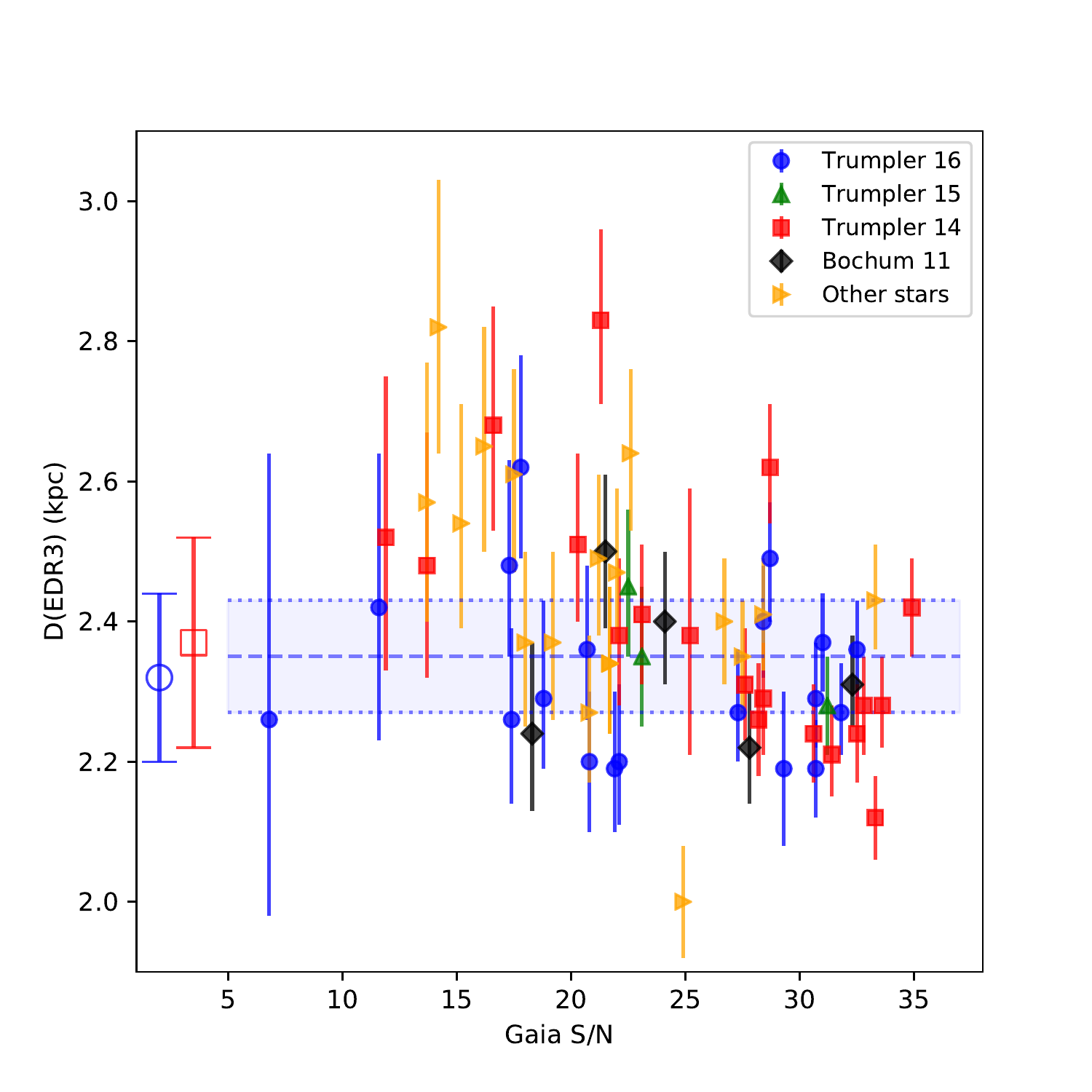}
\caption{Offset-corrected parallax distances from Gaia-EDR3 versus the significance
level, S/N  $= \varpi/\sigma_{\varpi}$, of the measurement.  We show individual values
for 46 OB-type stars in the Carina Nebula, color-coded for four clusters (Tr~16, Tr~15,
Tr~14, Bo~11) and other stars around Col 228.  The mean distance and (rms) variance
($2.35\pm0.08$~kpc) of the nine earliest O-type stars are shown by horizontal dashed 
lines and blue wash.  Mean distances and variances for Tr~16  (blue circle) and 
Tr~14 (red square) from Table 2 are  shown along the left side.  Several outliers with 
S/N $> 20$ could be foreground and background stars, and possibly escaping stars.   
These include three stars in Tr~14 (Tr~14-4 at 2.62~kpc, Tr~14-5 at 2.83~kpc, 
HD~303312 at 2.12~kpc) and HD~305518 at 2.00~kpc (in Col~228).  However, none
of these stars has an aberrant proper motion relative to cluster mean values. }
\end{figure}
 


\begin{figure}[!ht]
\includegraphics[angle=0,scale=1.05]{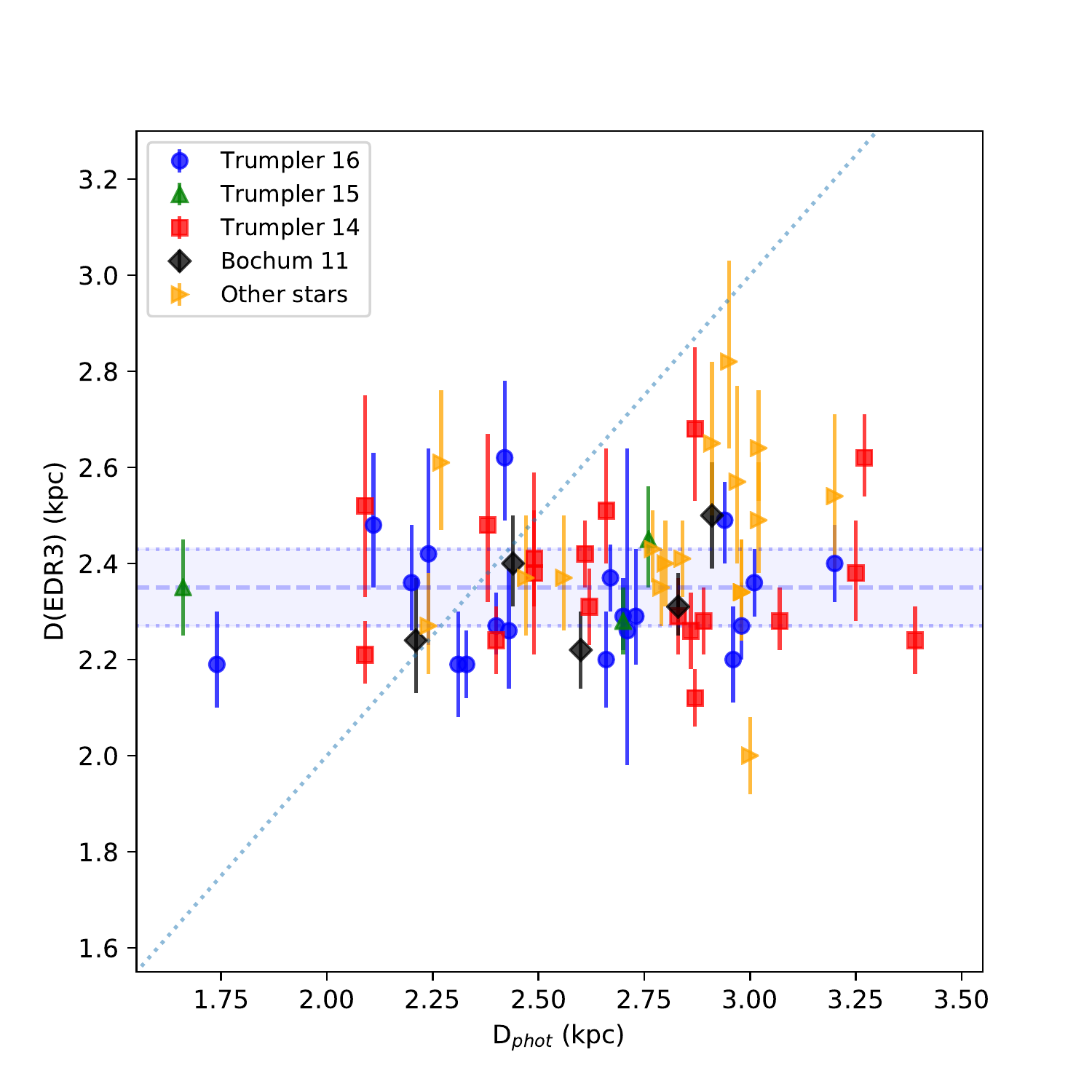}
\caption{Comparison of Gaia-EDR3 distance with spectrophotometric distances in Table 2, 
color-coded for four clusters (Tr~16, Tr~15, Tr~14, Bo~11) and other stars.  Horizontal lines 
and blue wash show the mean distance and rms variance to Car~OB1,
$D_{\rm EDR3} = 2.35 \pm 0.08$~kpc.   Two stars (Tr~14-5 and HD~93146B) with large,
but uncertain values $D_{\rm phot} > 4$~kpc, are not shown.  HD~93146A has a realistic 
photometric distance, $D_{\rm phot} = 2.30$~kpc, differing from that of HD~93146B.  
Using the $M_V$ scale of Martins \etal\ (2005) would bring these distances into somewhat 
better agreement, still with some anomalies  (see Tables 2 and 3).  }  
 
\end{figure}
 

\clearpage


\begin{deluxetable} {llll}
\tablecolumns{4}
\tabletypesize{\footnotesize}
\tablenum{1}
\tablewidth{0pt}
\tablecaption{Previous Distance Estimates\tablenotemark{a}}

\tablehead{
\colhead{Paper Name}
&\colhead{Trumpler 14}
&\colhead{Trumpler 16} 
&\colhead{Notes\tablenotemark{b}}
\\
\colhead{}
& \colhead{$D$ (kpc)}
&\colhead {$D$ (kpc)}
&\colhead{}  
}
   
\startdata
Th\'e \& Vleeming (1971)                & $2.0\pm0.2$                   & $2.5\pm0.2$               &  $R_V =  3.0$  \\
Th\'e \etal\ (1980)                             & 2.8 (2.3)                           & 2.8 (2.4)                       &  $R_V =  3.3$ (4.0)  \\
Walborn (1973a)                              & 3.5                                    & 2.6                                & UBV    \\
Feinstein \etal\ (1973)                     & $2.65 \pm10\%$            & $2.65 \pm10\%$        & UBV ($R_V = 4.0$) \\
Humphreys (1978)                          & $3.60\pm0.15$               & $2.69\pm0.15$           & Blue/Red Supergiants        \\
Turner \& Moffat (1980)                   & $2.70\pm0.17$              & $2.70\pm0.17$           & UBV ($R_V =  3.20\pm0.28$)  \\
Th\'e \etal\ (1980)                             & $2.3\pm0.3$                   & $2.4\pm0.3$                & UBV  ($R =  4$)   \\         
Levato \& Malaroda (1981)            & $\dots$                            & 2.6                                 & V (Col 228) ($R_V =$3.2-4.0)  \\    
Walborn (1982b)                              & 2.8                                    & 2.8                                & UBV  ($R_V = 3.0$)  \\                                        
Tapia \etal\ (1988)                           & $2.4\pm0.2$                   & $2.4\pm0.2$                & JHKL    \\
Morrell \etal\ (1988)                         & 2.8 (3.45)                        &  $\dots$                        &  V;   $R_V$ variable (3.2) \\
Massey \& Johnson (1993)            & $3.2 \pm4$\%                & $3.2 \pm4$\%              &  UBV ($R_V = 3.2$)    \\ 
Allen \& Hiller (1993)                       & $\dots$                            & $2.2\pm0.2$                & $\eta$ Car expansion \\
Walborn (1982b)                              & 2.8                                   &  2.8                                & UBV  ($R_V = 3.0$)  \\     
V\'azquez \etal\ (1996)                    & $3.2 \pm10$\%              & $\dots$                          & UBVRI     \\
Davidson \& Humphreys (1997)   &  $\dots$                            & $2.3\pm0.2$                & $\eta$ Car   \\ 
Meaburn (1999)                               &  $\dots$                            & $2.3\pm0.3$                & $\eta$ Car expansion \\
Davidson \etal\ (2001)                    &  $\dots$                            & $2.25\pm0.18$            & $\eta$ Car expansion \\
DeGioia-Eastwood \etal\ (2001)   & $3.6 \pm5\%$                 & $3.61 \pm5\%$            & UBV   \\
Tapia \etal\ (2003)                           & $2.8\pm0.2$                   & $2.5\pm0.2$                & UBVRIJHK    \\
Carraro \etal\ (2004)                       & $2.5\pm0.3$                    & $4.0\pm0.5$                & UBVRI       \\
Smith (2006a)                                  & $\dots$                              & $2.35\pm0.05$           & $\eta$ Car expansion \\
Hur \etal\ (2012)                              & $2.9\pm0.3$                     & $2.9\pm0.3$                & UBVI   \\
Davidson \etal\ (2018)                   & $3.0\pm0.2$                     & $2.6\pm0.2$                & Gaia-DR2       \\
Kuhn \etal\ (2019)                           & $2.64^{+0.31}_{-0.25}$ & $2.61^{+0.31}_{-0.25}$  & Gaia-DR2  \\              
Povich \etal\ (2019)                        & $2.50^{+0.28}_{-0.23}$  & $2.50^{+0.28}_{-0.23}$ & Gaia-DR2  \\
                                               
\enddata

\tablenotetext{a} {Photometric distance estimates were estimated with various assumptions about 
the ratio of total-to-selective extinction, $R_V = A_V / E(B-V)$.  These extinction corrections to $V$ 
magnitudes provided distance moduli DM = $V - M_V - A_V$, typically ranging from 11.8 to 13.0 
and converted to physical distances by the relation $D_{\rm phot} = (10~{\rm pc}) 10^{{\rm DM}/5}$. }

\tablenotetext{b} {Notes on photometry methods and assumed values of $R_V$ (between 3.0 and 4.2).
In a review, Walborn (2012) estimated a Tr 16 distance of $(2.25~{\rm kpc})(0.8)^{(R_V - 4)}$. }

\end{deluxetable}



\begin{deluxetable}{lll llc cll}
\tablecolumns{9}
\tabletypesize{\scriptsize}
\tablenum{2}
\tablewidth{0pt}
\tablecaption{Gaia Parallaxes and Distances\tablenotemark{a} }

\tablehead{
\colhead{Star Name}
&\colhead{Other}
&\colhead{SpT} 
&\colhead{EDR3 ($\varpi \pm \sigma_{\varpi}$)}
&\colhead{Offset}
&\colhead{PM} 
&\colhead{$D_{\rm EDR3}$}
&\colhead{$D_{\rm phot}$} 
&\colhead{Notes}
\\
\colhead{}
& \colhead{}
& \colhead{}     
& \colhead{(mas)}
& \colhead{(mas)}
& \colhead{(mas/yr)} 
& \colhead{(kpc)} 
& \colhead{(kpc)} 
& \colhead{(see text)}
}

\startdata
       
{\bf Trumpler 16:}                       &                    &                 &                                      &            &             &                             &         &         \\        
                                                      &                    &                 &                                      &            &             &                             &         &         \\                                                                                                                               
{HD  93204 (\#44)}                    & Tr 16-178  & O5.5 V   & $0.4240\pm0.0226$ & 0.012 & $7.524\pm0.025$ & 2.29[2.19,2.43] & 2.73/2.40 &     \\ 
{HD  93205 (\#45)}                    & V560 Car  & O3.5 V   & $0.4308\pm0.0248$ & 0.012 & $8.169\pm0.028$ & 2.26[2.14,2.39] & 2.43/2.34 & +O8 V  \\
{HD  93250 (\#48)}                    & Tr 16-101  & O4 III     & $0.4115\pm0.0199$ & 0.012 & $7.754\pm0.023$ & 2.36[2.26,2.48] & 2.20/1.99 &      \\
{HD 303308 (\#137)}                & Tr 16-7       & O4.5 V  & $0.4432\pm0.0213$ & 0.012 & $7.033\pm0.023$ & 2.20[2.10,2.30] & 2.66/2.40 &     \\
{CPD-$59^{\circ}2600$ (\#3)} & Tr 16-100  & O6 V      & $0.3914\pm0.0226$ & 0.012 & $7.362\pm0.026$ & 2.48[2.35,2.63] & 2.11/1.86 &      \\             
{CPD-$59^{\circ}2603$ (\#4)} & V572 Car  & O7.5 V   & $0.3690\pm0.0207$ & 0.012 & $7.529\pm0.024$ & 2.62[2.49,2.78] & 2.42/2.14 & 3, +B0 V     \\            
{CPD-$59^{\circ}2591$}          & Tr 16-21     & O8.5 V   & $0.4056\pm0.0131$ & 0.017 & $7.178\pm0.016$ & 2.37[2.30,2.44] & 2.67/2.37 &     \\   
{CPD-$59^{\circ}2624$}          & Tr 16-9       & O9.7 IV  & $0.4302\pm0.0631$ & 0.012 & $7.274\pm0.079$ & 2.26[1.98,2.64] & 2.71/2.61 & 1, 3 (not 2634)  \\               
{CPD-$59^{\circ}2626$}          & Tr 16-23     & O7.5 V   & $0.4009\pm0.0346$ & 0.013 & $7.768\pm0.041$ & 2.42[2.23,2.64] & 2.24/1.98 &    \\                 
{CPD-$59^{\circ}2627$}          & Tr 16-3       & O9.5 V   & $0.3904\pm0.0136$ & 0.012 & $7.397\pm0.016$ & 2.49[2.40,2.57] & 2.94/2.62 &     \\                  
{CPD-$59^{\circ}2628$}          & V573 Car   & O9.5 V  & $0.4428\pm0.0200$ & 0.011 & $7.537\pm0.022$ & 2.20[2.11,2.31] & 2.96/2.64 & +B0.5 V  \\                 
{CPD-$59^{\circ}2629$}          & Tr 16-22     & O8.5 V  & $0.4259\pm0.0134$ & 0.014 & $7.145\pm0.016$ & 2.27[2.21,2.34] & 2.40/2.13 &     \\   
{V662 Car}                                 &                      & O5 V     & $0.4126\pm0.0127$ & 0.011 & $7.195\pm0.014$ & 2.36[2.29,2.43] & 3.01/2.64 &  + B0 V \\             
{CPD-$59^{\circ}2635$}          & Tr 16-34    & O8 V      & $0.5171\pm0.0186$ & 0.012 & $7.258\pm0.021$ & 1.89[1.82,1.96] & 2.50/2.23 & +O9.5 V  \\        
{CPD-$59^{\circ}2636$AB}     & Tr 16-110  & O8 V      & not available             &$\dots$&$\dots$                     & not available     & 2.34/2.08 & 3, +O8 V  \\                                      
{CPD-$59^{\circ}2641$}          & Tr 16-112  & O6 V      & $0.4445\pm0.0145$ & 0.012 & $7.522\pm0.017$ & 2.19[2.12,2.26] & 2.33/2.05 &     \\          
{CPD-$59^{\circ}2644$}          & Tr 16-115  & O9 V      & $0.4286\pm0.0157$ & 0.011 & $7.258\pm0.018$ & 2.27[2.20,2.36] & 2.98/2.66 &    \\    
{HD 303316}                             &                     & O7 V      & $0.4241\pm0.0138$ & 0.012 & $6.732\pm0.016$ & 2.29[2.22,2.37] & 2.70/2.39 &   \\
{ALS 15210}                              & Tr 16-244  & O3.5 If*  & $0.4008\pm0.0141$ & 0.016 & $7.760\pm0.016$ & 2.40[2.32,2.48] & 2.48/2.53 &    \\
{HD 93162}                                & WR25        & O2.5 If*  & $0.4450\pm0.0203$ & 0.011 & $7.450\pm0.023$ & 2.19[2.10,2.30] &1.74/1.35 & WN6+OB  \\
{HD 93343}                                &                    & O8 V      & $0.4452\pm0.0228$ & 0.012 & $7.338\pm0.018$ & 2.19[2.08,2.30] & 2.31/2.05 &  \\
 {\bf  Mean values\tablenotemark{b}}  &      &                 &                                &   & ${\bf 7.417\pm0.327}$ & ${\bf 2.32\pm0.12}$ & {\bf 2.56/2.31} &    \\                                                                             
                                                     &                    &                 &                                      &             &          &           &      &         \\      
                                                     &                    &                 &                                      &             &          &           &      &         \\                                                                        
{\bf Trumpler 14:}                      &                    &                 &                                      &             &          &           &       &         \\   
                                                     &                    &                 &                                      &             &          &           &       &         \\                                                                                                                                                                                                 
{HD 93129A (\#42)}                 &                    & O2 I         & $0.4037\pm0.0175$ & 0.012 & $6.874\pm0.019$ & 2.41[2.31,2.51] & 2.49         & 3, AaAb   \\
{HD 93129B}                            &                     & O3.5 V    & $0.4073\pm0.0162$ & 0.012 & $6.969\pm0.020$ & 2.38[2.30,2.48] & 2.49         & 3   \\
{HD 93128}                               &                     & O3.5 V   & $0.4256\pm0.0150$ & 0.012 & $6.952\pm0.016$ & 2.29[2.21,2.37] & 2.83/2.73 &      \\
{HD 93160}                               &                     & O7 III      & $0.3845\pm0.0323$ & 0.012 & $7.062\pm0.034$ & 2.52[2.33,2.75] & 2.09/1.94 &       \\
{HD 93161A}                            & Tr 14-176  & O7.5 V   & $0.3399\pm0.0281$ & 0.012 & $6.842\pm0.029$ & 2.84[2.69,3.09] & 2.28/2.04 & 3, +O9 V   \\
{HD 93161B}                            & Tr 14-176  & O6.5 IV  & $0.3859\pm0.0282$ & 0.017 & $7.137\pm0.029$ & 2.48[2.32,2.67] & 2.41/2.18 & 3    \\
{HD 303311}                            &                     & O6 V       & $0.4343\pm0.0142$ & 0.012 & $6.550\pm0.016$ & 2.24[2.17,2.31] & 3.39/2.95 & 3     \\
{HD 303312}                            & V725 Car  & O9.7 IV    & $0.4599\pm0.0138$ & 0.012 & $6.819\pm0.015$ & 2.12[2.06,2.18] & 2.87/2.76 &       \\
{CPD-$58^{\circ}2611$}        & Tr 14-20     & O6 V       & $0.4263\pm0.0127$ & 0.012 & $6.992\pm0.014$ & 2.28[2.22,2.35] & 3.07/2.70 &        \\
{CPD-$58^{\circ}2620$}        & Tr 14-8       & O7 V       & $0.3610\pm0.0217$ & 0.012 & $7.172\pm0.024$ & 2.68[2.53,2.85] & 2.87/2.54 &       \\                  
{CPD-$58^{\circ}2627$}        &                     & O9.5 V    & $0.4027\pm0.0182$ & 0.017 & $6.986\pm0.019$ & 2.38[2.28,2.49] & 3.25/2.90 &      \\               
{Tr 14-3}                                    &                    & B0.5 V     & $0.4018\pm0.0115$ & 0.012 & $6.765\pm0.012$ & 2.42[2.35,2.49] & 2.61 & 2, 3  \\
{Tr 14-4}                                    &                    & B0 V        & $0.3647\pm0.0127$ & 0.017 & $7.033\pm0.015$ & 2.62[2.54,2.71] & 3.27 & 2, 3  \\
{Tr 14-5}                                    &                    & O9 V        & $0.3272\pm0.0154$ & 0.026 & $6.995\pm0.019$ & 2.83[2.71,2.96] & 4.06/3.61 & 2, 3  \\
{Tr 14-6}                                    &                    & B1 V        & $0.4199\pm0.0149$ & 0.023 & $6.751\pm0.017$ & 2.26[2.18,2.34] & 2.86 & 2, 3    \\
{Tr 14-9}                                    &                    & O8.5 V    & $0.4359\pm0.0134$ & 0.011 & $6.835\pm0.015$ & 2.24[2.17,2.31] & 2.40/2.13 &          \\
{Tr 14-27}                                 &                     & B1.5 V    & $0.4270\pm0.0136$ & 0.025 & $6.567\pm0.015$ & 2.21[2.15,2.28] & 2.09 & 2, 3   \\
{ALS 15204}                            &                     & O7.5 V    & $0.4198\pm0.0152$ & 0.014 & $7.215\pm0.016$ & 2.31[2.23,2.39] & 2.62/2.31 &         \\
{ALS 15206}                            & CPD-$58^{\circ}2625$ & O9.2 V    & $0.3854\pm0.0190$ & 0.013 & $6.863\pm0.021$ & 2.51[2.40,2.64] & 2.66/2.37 &        \\
{ALS 15207}                            &  Tr 14-21   & O9 V       & $0.4270\pm0.0130$ & 0.012 & $7.507\pm0.014$ & 2.28[2.21,2.35] & 2.89/2.58 &        \\
{\bf Mean values\tablenotemark{b}}   &      &                 &                              &     & ${\bf 6.944\pm0.221}$ & ${\bf 2.37\pm0.15}$ & {\bf 2.71/2.47} &   \\                                                               
                                                   &                    &                 &                                       &             &          &           &       &  \\       
                                                   &                    &                 &                                       &             &          &           &       &  \\      
                                                   &                    &                 &                                       &             &          &           &       &  \\      
                                                  &                     &                 &                                       &             &          &           &       &   \\      
                                                   &                    &                 &                                       &             &          &           &       &  \\          
{\bf Trumpler 15:}                    &                    &                 &                                       &             &          &            &       &  \\   
                                                   &                    &                 &                                      &              &          &            &       &         \\                                                                                                                                                                             
{HD 93249}                              &                   & O9 III       & $0.3966\pm0.0176$ & 0.012 & $6.343\pm0.019$ & 2.45[2.35,2.56] & 2.76/2.70 &  +O3 III  \\ 
{HD 303304}                            &                   & O7 V       & $0.4270\pm0.0137$ & 0.012 & $7.063\pm0.017$ & 2.28[2.21,2.35] & 2.70/2.39 &  2, 3 \\ 
{HD 93190}                              &                   & O9.7 V:   & $0.4141\pm0.0179$ & 0.012 & $6.915\pm0.020$ & 2.35[2.25,2.45] & 1.66/1.54 &  2, 3   \\
{\bf Mean values\tablenotemark{b}}  &     &                 &                           &      & ${\bf 6.774\pm 0.380}$ & ${\bf 2.36\pm0.09}$  & {\bf 2.37/2.21} &             \\      
                                                    &                  &                 &                                       &            &          &           &        &      \\        
                                                    &                  &                 &                                       &            &          &           &        &     \\        
{\bf Bochum 11:}                      &                   &                &                                       &            &          &           &        &       \\   
                                                    &                   &                &                                       &            &        &            &         &         \\             
{HD 93576}                              & Bo 11        & O9.5 IV  & $0.3887\pm0.0181$ & 0.012 & $6.488\pm0.020$ & 2.50[2.39,2.61] & 2.91/2.70 &    \\
{HD 93632}                              & Bo 11        & O5 If       & $0.4049\pm0.0168$ & 0.012 & $6.723\pm0.021$ & 2.40[2.31,2.50] & 2.44/2.49 &    \\
{HD 305539}                           & Bo 11         & O8 V      & $0.4204\pm0.0130$ & 0.012 & $6.627\pm0.014$ & 2.31[2.25,2.38] & 2.83/2.51 &   \\
{HD 305612}                           & Bo 11         & O8 V      & $0.4390\pm0.0158$ & 0.012 & $6.511\pm0.017$ & 2.22[2.14,2.30] & 2.60/2.30 &    \\
{ALS 18083}                            & Bo 11         & O9.7 V  & $0.4334\pm0.0237$ & 0.013 & $6.609\pm0.027$ & 2.24[2.13,2.37] & 2.21/2.05 &   \\
{\bf Mean values\tablenotemark{b}} &       &                &                           &   & ${\bf 6.592\pm0.095}$ & ${\bf 2.33\pm0.12}$ & {\bf 2.60/2.41} &   \\     
                                                  &                     &                &                                       &                                      &         \\                                                                                                                                                                                       
                                                  &                     &                &                                       &                                      &         \\                                                                                                                                                                                           
{\bf  Other Stars:}                    &                    &                 &                                       &                                       &         \\     
                                                  &                    &                 &                                       &                                       &                                      &        \\                                                                                                                                                                                                                                       
 {HD 93028 (\#41)}                & Col 228     & O9 IV     & $0.3771\pm0.0275$ & 0.012 & $6.651\pm0.031$ & 2.57[2.40,2.77] & 2.97/2.74 &      \\
{HD 93222 (\#47)}                 & Col 228     & O7 III      & $0.4102\pm0.0228$ & 0.012 & $6.832\pm0.025$ & 2.37[2.25,2.50] & 2.47/2.30 &  3  \\
{HD 93146A (\#43)}              & Col 228     & O7 V       & $0.3288\pm0.0358$ & 0.012 & $6.807\pm0.040$ & 2.93[2.66,3.28] & 2.30/2.03 &   3  \\
{HD 93146B}                          & Col 228     & O9.7 IV  & $0.3937\pm0.0179$ & 0.011 & $7.058\pm0.019$ & 2.47[2.37,2.59] & 4.29/4.13 &  3  \\
{HD 93206A (\#46)}              & QZ Car      & O9.7 Ib   & $0.7356\pm0.1086$ & 0.013 & $6.252\pm0.125$ & 1.34[1.17,1.56] & 2.23/2.33 &  3, +O8 III   \\
{HD 93130}                            & Col 228     & O6.5 III   & $0.3710\pm0.0212$ & 0.012 & $7.215\pm0.024$ & 2.61[2.47,2.76] & 2.27/2.09 &    \\
{HD 93027}                            & Col 228     & O9.5 IV   & $0.3424\pm0.0242$ & 0.012 & $7.377\pm0.030$ & 2.82[2.64,3.03] & 2.95/2.74 &    \\
{CPD-$59^{\circ}2554$}      & Col 228     & O9.5 IV   & $0.3645\pm0.0226$ & 0.011 & $7.024\pm0.025$ & 2.66[2.51,2.83] & 2.91/2.70 &    \\
{CPD-$59^{\circ}2551$}      &  Col 228    & O9 V       & $0.3901\pm0.0184$ & 0.011 & $6.999\pm0.020$ & 2.49[2.38,2.61] & 3.02/2.70 &    \\
{CPD-$59^{\circ}2610$}      &  Col 228     & O8.5 V   & $0.4158\pm0.0192$ & 0.011 & $7.097\pm0.023$ & 2.34[2.24,2.45] & 2.98/2.64 &    \\
{CPD-$59^{\circ}2673$}      & Col 228      & O5.5 V   & $0.3657\pm0.0162$ & 0.013 & $6.743\pm0.019$ & 2.64[2.53,2.76] & 3.02/2.65 &   \\
{HD 305438}                          & Col 228      & O8 V      & $0.3820\pm0.0252$ & 0.012 & $8.195\pm0.030$ & 2.54[2.39,2.71] & 3.20/2.84 &   \\
{HD 305518}                          & Col 228     & O9.7 III   & $0.4885\pm0.0196$ & 0.012 & $7.273\pm0.021$ & 2.00[1.92,2.08] & 3.00/3.05 &    \\
{HD 305523}                          & Col 228     & O9 II-III   & $0.4099\pm0.0214$ & 0.012 & $6.194\pm0.025$ & 2.37[2.26,2.50] & 2.56/2.47 &    \\
{HD 305524}                          & Col 228     & O6.5 V   & $0.4055\pm0.0152$ & 0.012 & $7.623\pm0.017$ & 2.40[2.31,2.49] & 2.80/2.46 &    \\
{HD 305532}                          & TrC             & O6.5 V   & $0.4031\pm0.0142$ & 0.012 & $7.070\pm0.016$ & 2.41[2.33,2.49] & 2.84/2.50 &   \\
{CPD-$59^{\circ}2661$}      & TrC             & O9.5 V   & $0.3992\pm0.0145$ & 0.026 & $7.017\pm0.017$ & 2.35[2.27,2.43] & 2.79/2.49 &    \\
{HD 305536}                          & Col 228     & O9.5 V   & $0.4296\pm0.0206$ & 0.011 & $7.275\pm0.023$ & 2.27[2.17,2.38] & 2.24/2.00 &    \\
{HD 305619}                          & Col 228     & O9.7 II   & $0.3992\pm0.0120$ & 0.012 & $7.208\pm0.014$ & 2.43[2.36,2.51] & 2.77/2.61 &     \\
{CPD-$59^{\circ}2610$}      &  Col 228     & O8.5 V  & $0.4158\pm0.0192$ & 0.011 & $7.097\pm0.023$ & 2.34[2.24,2.45] & 2.98/2.64 &    \\
{\bf Mean values\tablenotemark{b}}  &      &               &        &        & ${\bf 7.080\pm0.207}$ & ${\bf 2.45\pm0.18}$ & {\bf 2.75/2.53} &   \\       

\enddata

\tablenotetext{a} {Parallax measurements for 68 OB~stars in Carina OB1 association, including 
associated clusters (Trumpler~14,  Trumpler~15, Trumpler~16, Collinder~228, Bochum~11, and
Treasure Chest).  Column~1 gives star name, with internal ID for 11 stars in the Shull \& Danforth 
(2019) survey.  Column~2 gives other names or cluster membership.  Column 3 gives  the SpT from 
GOSS except for seven stars.  Column 4 lists parallaxes and errors from Gaia-EDR3.   
Column~5 gives the DR3 parallax offset (Lindegren \etal\ 2021) used for distance $D_{\rm EDR3}$ 
(and 1$\sigma$ range) in column 7.  Column 6 is the total proper motion (PM).  
Column 8 gives two photometric distances, based on GOSS photometry and SpTs when available,
and absolute magnitudes $M_V$ from Bowen \etal\ (2008) and Martins \etal\ (2005) respectively.  
Column 9 provides notes on binary companions and also:  
(1) GOSS typo in the identification of CPD-$59^{\circ}2624$; 
(2) seven stars not found in GOSS; 
(3) methods for deriving photometric distances (see Appendix A).}

\tablenotetext{b} {Mean EDR3 distances and (rms) variances of distance distributions were computed, 
omitting six stars with uncertain or missing data:  two each in Tr~14, Tr~16, and Col~228 (see Section~2).   
Mean values for proper motion also omitted discrepant outliers (see Section 3).  
The mean photometric distances are 10--14\% higher: $2.56\pm0.28$~kpc (Tr~16); $2.71\pm0.36$~kpc 
(Tr~14);  $2.37\pm0.61$~kpc (Tr~15); $2.60\pm0.29$~kpc (Bo~11); and $2.75\pm0.33$~kpc (Other Stars).  
 }

\end{deluxetable}



\begin{deluxetable}{llc ccl}
\tablecolumns{6}
\tabletypesize{\scriptsize}
\tablenum{3}
\tablewidth{0pt}
\tablecaption{Stellar Distance Data\tablenotemark{a} }

\tablehead{
 \colhead{SpT} 
& \colhead{Star Name}
& \colhead{$\Delta M_V$} 
&\colhead{$D_{\rm phot}^{(1)}$} 
&\colhead{$D_{\rm phot}^{(2)}$} 
&\colhead{$D_{\rm EDR3}$}
\\
   \colhead{}
& \colhead{}   
& \colhead{(mag)} 
& \colhead{(kpc)} 
& \colhead{(kpc)} 
& \colhead{(kpc)} 
}

\startdata

O3.5 V    &  {HD  93205}                      & 0.08          & 2.43 & 2.34 &  2.26(2.14,2.39)    \\    
O3.5 V    &  {HD 93128}                       & 0.08          & 2.83 & 2.73 &  2.29(2.21,2.37)    \\
O4.5 V    &  {HD 303308}                     & 0.22          & 2.66 & 2.40 & 2.20(2.10,2.30)     \\
O5 V       &  {V662 Car}                         & 0.29          & 3.01 & 2.64 & 2.36(2.29,2.43)     \\       
O5.5 V    & {HD  93204}                       & 0.28           & 2.73 & 2.40 & 2.29(2.19,2.43)      \\ 
O5.5 V    & {CPD-$59^{\circ}2673$}  & 0.28           & 3.02 & 2.65 & 2.64(2.53,2.76)      \\ 
O6 V       & {CPD-$59^{\circ}2600$}  & 0.28           & 2.11 &1.86  & 2.48(2.35,2.63)      \\             
O6 V       & {CPD-$59^{\circ}2641$}  & 0.28           & 2.33 & 2.05 & 2.19(2.12,2.26)      \\
O6 V       & {CPD-$58^{\circ}2611$}  & 0.28           & 3.07 & 2.70 & 2.28(2.22,2.35)      \\
O6 V       & {HD 303311}                      & 0.28           & 3.39 & 2.99 & 2.24(2.17,2.31)     \\    
O6.5 V    & {HD 305524}                      & 0.28           & 2.80 & 2.46 & 2.40(2.31,2.49)      \\
O6.5 V    & {HD 305532}                      & 0.28           & 2.84 & 2.50 & 2.41(2.33,2.49)    \\
O7 V       & {HD 303316A}                   & 0.27           & 2.70 & 2.39 & 2.29(2.22,2.37)    \\
O7 V       & {CPD-$58^{\circ}2620$}  & 0.27           & 2.87 & 2.54 & 2.68(2.53,2.85)    \\ 
O7.5 V   & {CPD-$59^{\circ}2603$}  & 0.27            & 2.42 & 2.14 & 2.62(2.49,2.78)    \\
O7.5 V   & {CPD-$59^{\circ}2626$}  & 0.27            & 2.24 & 1.98 & 2.42(2.23,2.64)   \\
O7.5 V   & {ALS 15204}                      & 0.27            & 2.62 & 2.31 & 2.31(2.23,2.39)   \\
O8 V      & {HD 93343}                        & 0.26            & 2.31 & 2.05 & 2.19(2.08,2.30)  \\
O8 V      & {HD 305539}                      & 0.26            & 2.83 & 2.51 & 2.31(2.25,2.38)  \\ 
O8 V      & {HD 305612}                      & 0.26            & 2.60 & 2.30 & 2.22(2.14,2.30)   \\ 
O8 V      & {HD 305438}                      & 0.26            & 3.20 & 2.84 & 2.54(2.39,2.71)   \\ 
O8.5 V   & {CPD-$59^{\circ}2591$}  & 0.26            & 2.67 & 2.37 & 2.37(2.30,2.44)    \\
O8.5 V   & {CPD-$59^{\circ}2629$}  & 0.26            & 2.40 & 2.13 & 2.27(2.21,2.34)    \\
O8.5 V   & {Tr 14-9}                              & 0.26            & 2.40 & 2.13 & 2.24(2.17,2.31)   \\ 
O8.5 V   & {CPD-$59^{\circ}2610$}   & 0.26           & 2.98 & 2.64 & 2.34(2.24,2.45)   \\ 
O9 V      & {CPD-$59^{\circ}2644$}   & 0.25           & 2.98 & 2.66 & 2.27(2.20,2.36)    \\
O9 V      & {ALS 15207}                       & 0.25           & 2.89 & 2.58 & 2.28(2.21,2.35)     \\
O9 V      & {CPD-$59^{\circ}2551$}   & 0.25           & 3.02 & 2.70 & 2.49(2.38,2.61)    \\ 
O9.2 V   & {ALS 15206}                       & 0.25           & 2.66 & 2.37 & 2.51(2.40,2.64)   \\
O9.5 V   & {CPD-$59^{\circ}2627$}  & 0.25           & 2.94 & 2.62 & 2.49(2.40,2.57)    \\
O9.5V    & {CPD-$59^{\circ}2628$}  & 0.25           & 2.96 & 2.64 & 2.20(2.11,2.31)    \\ 
O9.5 V   & {CPD-$58^{\circ}2627$}  & 0.25           & 3.25 & 2.90 & 2.38(2.28,2.49)   \\
O9.5 V   & {HD 305536}                      & 0.25           & 2.24 & 2.00 & 2.27(2.17,2.38)   \\ 
O9.7 V   & {ALS 18083}                      & 0.16            & 2.21 & 2.05 & 2.24(2.13,2.37)    \\        
B0 V       & {Tr 14-4}                             &                     & 3.27 &          & 2.62(2.54,2.71)   \\
B0.5 V    & {Tr 14-3}                             &                     & 2.61 &          & 2.42(2.35,2.49)    \\
B1 V       & {Tr 14-6}                             &                     & 2.86 &          & 2.26(2.18,2.34)   \\                              
B1.5 V    & {Tr 14-27}                           &                    & 2.09 &          & 2.21(2.15,2.28)    \\   
                                                                                                                                                                                                                                                                                
\enddata

\tablenotetext{a} {Spectrophotometric and parallax distances for 39 main-sequence stars in 
Car~OB1 from O3.5~V to B1.5~V.  There were no O4~V stars in the sample.     
$D_{\rm phot}^{(1)}$ and $D_{\rm phot}^{(2)}$ correspond to absolute magnitudes $M_V$
from Bowen \etal\ (2008) and Martins \etal\ (2005), respectively, and all with GOSS photometry 
and SpTs.  Columns~1 and 2 give the SpT and star name.  
Column 3 gives the difference $\Delta M_V$ in absolute magnitudes (the Martins magnitudes 
are fainter, but none were provided for B-type stars).  
The Gaia-EDR3 distances $D_{\rm EDR3}$ and range $(D_{\rm min}, D_{\rm max})$ are 
based on offset-corrected parallaxes  and $1\sigma$ errors. }

\end{deluxetable}



\begin{deluxetable}{lllllcc}
\tablecolumns{7}
\tabletypesize{\scriptsize}
\tablenum{4}
\tablewidth{0pt}
\tablecaption{Gaia Proper Motions\tablenotemark{a} }

\tablehead{
\colhead{Star Name}
&\colhead{PM$_{\rm RA}^{\rm (Sun)}$}
&\colhead{PM$_{\rm Dec}^{\rm (Sun)}$} 
&\colhead{PM$_{\rm RA}^{\rm (Car)}$}
&\colhead{PM$_{\rm Dec}^{\rm (Car)}$} 
&\colhead{PM$_{\rm tot}^{\rm (Car)}$} 
&\colhead{V$_{\rm tran}$} 
\\
\colhead{}
& \colhead{(mas/yr)} 
& \colhead{(mas/yr)} 
& \colhead{(mas/yr)} 
& \colhead{(mas/yr)} 
& \colhead{(mas/yr)} 
& \colhead{(km/s)} 
}

\startdata                                                        
{\bf Trumpler 14:}                      &                                   &                                  &                    &                   &         \\                                                                                                                                                                                  
{HD 93129A}                             & $-6.277\pm0.019$ & $2.801\pm0.019$ & $+0.303$ & $+0.616$ & $0.686\pm0.082$ & $7.64\pm0.91$  \\  
{HD 93129B}                            & $-6.618\pm0.020$ & $2.182\pm0.019$ & $-0.038$   &  $-0.003$ & $0.038\pm0.063$ & $0.42\pm0.70$   \\
{HD 93128}                               & $-6.616\pm0.016$ & $2.135\pm0.016$ & $-0.036$   & $-0.050$  & $0.062\pm0.078$ & $0.69\pm0.87$  \\
{HD 93160}                               & $-6.389\pm0.034$ & $3.011\pm0.034$ &  $+0.191$ & $+0.826$ & $0.848\pm0.090$ & $9.45\pm1.00$   \\
{HD 93161A}                            & $-6.509\pm0.029$ & $2.109\pm0.029$ &  $+0.071$ & $-0.076$  & $0.104\pm0.079$ & $1.16\pm0.88$   \\
{HD 93161B}                            & $-6.890\pm0.029$ & $1.862\pm0.030$ & $-0.310$   & $-0.323$  & $0.448\pm0.079$ & $4.99\pm0.89$  \\
{HD 303311}                            & $-6.213\pm0.016$ & $2.074\pm0.016$ & $+0.367$  & $-0.111$  & $0.383\pm0.065$ & $3.27\pm0.72$   \\
{HD 303312}                            & $-6.511\pm0.015$ & $2.025\pm0.015$ & $+0.069$  & $-0.160$  & $0.174\pm0.082$ & $1.94\pm0.92$    \\
{CPD-$58^{\circ}2611$}        & $-6.747\pm0.014$ & $1.837\pm0.014$ & $-0.167$   & $-0.348$  & $0.386\pm0.081$ & $4.30\pm0.01$   \\
{CPD-$58^{\circ}2620$}        & $-6.981\pm0.024$ & $1.642\pm0.023$ & $-0.401$   & $-0.543$  & $0.675\pm0.080$ & $7.52\pm0.89$   \\                  
{CPD-$58^{\circ}2627$}        & $-6.483\pm0.019$ & $2.603\pm0.019$ & $+0.097$   & $+0.418$ & $0.429\pm0.085$& $4.78\pm0.95$   \\               
{Tr 14-3}                                    & $-6.450\pm0.012$ & $2.039\pm0.014$ & $+0.130$  & $-0.146$  & $0.195\pm0.076$ & $2.17\pm0.84$   \\
{Tr 14-4}                                    & $-6.704\pm0.015$ & $2.126\pm0.014$ & $-0.124$  & $-0.059$   & $0.137\pm0.067$ & $1.53\pm0.75$    \\
{Tr 14-5}                                    & $-6.619\pm0.019$ & $2.261\pm0.016$ & $-0.039$  & $+0.076$  & $0.085\pm0.082$ & $0.95\pm0.91$    \\
{Tr 14-6}                                    & $-6.444\pm0.017$ & $2.014\pm0.015$ & $+0.136$ & $-0.171$   & $0.218\pm0.077$ & $2.43\pm0.86$   \\
{Tr 14-9}                                    & $-6.486\pm0.015$ & $2.156\pm0.014$ & $+0.094$ & $-0.029$   & $0.098\pm0.064$ & $1.09\pm0.72$   \\
{Tr 14-27}                                 & $-6.298\pm0.015$ & $1.859\pm0.017$ & $+0.282$  & $-0.326$  & $0.431\pm0.076$ & $4.80\pm0.85$    \\
{ALS 15204}                            & $-6.702\pm0.016$ & $2.670\pm0.016$ & $-0.122$  & $+0.485$  & $0.500\pm0.083$ & $5.57\pm0.92$    \\
{ALS 15206}                            & $-6.369\pm0.021$ & $2.557\pm0.021$ & $+0.211$ & $+0.372$  & $0.428\pm0.082$ & $4.77\pm0.91$    \\
{ALS 15207}                            & $-7.302\pm0.014$ & $1.741\pm0.014$ & $-0.722$  & $-0.444$  & $0.848\pm0.069$  & $9.45\pm0.77$    \\
 {\bf  Mean values}                  & ${\bf -6.580\pm 0.060}$ & ${\bf 2.185\pm 0.084}$ &             &       & ${\bf 0.359\pm0.058}$ & ${\bf 4.00\pm0.65}$  \\                       
                                                   &                                   &                                  &                   &                   &                                     &                               \\                                                                                                                                 
{\bf Trumpler 16:}                    &                                   &                                  &                   &                   &                                     &                               \\                                                                                                                                                             
{HD  93204}                             & $-7.089\pm0.025$ & $2.523\pm0.026$ & $-0.158$ & $-0.089$  & $0.181\pm0.067$ & $2.02 \pm0.75$   \\ 
{HD  93205}                             & $-7.619\pm0.028$ & $2.947\pm0.027$ & $-0.688$ & $+0.335$ & $0.765\pm0.068$ & $8.52 \pm0.76$    \\
{HD  93250}                             & $-7.116\pm0.023$ & $3.081\pm0.023$ & $-0.185$ & $+0.469$ & $0.504\pm0.063$ & $5.61 \pm0.70$    \\
{HD 303308}                            & $-6.642\pm0.023$ & $2.313\pm0.023$ & $+0.289$ & $-0.299$& $0.416\pm0.065$ & $5.12 \pm0.72$     \\
{CPD-$59^{\circ}2600$}        & $-6.970\pm0.026$ & $2.368\pm0.026$ & $-0.039$ & $-0.244$ & $0.247\pm0.064$ & $2.75 \pm0.71$     \\             
{CPD-$59^{\circ}2603$}        & $-6.971\pm0.024$ & $2.844\pm0.023$ & $-0.040$ & $+0.232$ & $0.235\pm0.063$ & $2.62 \pm0.70$     \\            
{CPD-$59^{\circ}2591$}        & $-6.707\pm0.016$ & $2.559\pm0.016$ & $-0.224$ & $-0.053$ & $0.230\pm0.065$ & $2.56 \pm0.72$     \\   
{CPD-$59^{\circ}2624$}        & $-6.900\pm0.079$ & $2.302\pm0.076$ & $+0.031$ & $-0.310$ & $0.312\pm0.096$ & $3.48 \pm1.06$    \\               
{CPD-$59^{\circ}2626$}        & $-7.304\pm0.041$ & $2.645\pm0.038$ & $-0.373$ & $+0.033$ & $0.374\pm0.075$ & $4.17 \pm0.84$     \\                 
{CPD-$59^{\circ}2627$}        & $-6.939\pm0.016$ & $2.564\pm0.016$ & $-0.008$ & $-0.048$  & $0.049\pm0.060$ & $0.55 \pm0.67$     \\                  
{CPD-$59^{\circ}2628$}        & $-6.998\pm0.021$ & $2.799\pm0.025$ & $-0.067$ & $+0.187$ & $0.199\pm0.063$ & $2.22 \pm0.71$      \\                 
{CPD-$59^{\circ}2629$}        & $-6.527\pm0.016$ & $2.906\pm0.014$ & $+0.404$ & $+0.294$ & $0.500\pm0.063$ & $5.57 \pm0.70$     \\   
{V662 Car}                               & $-6.724\pm0.014$ & $2.561\pm0.013$ & $+0.207$ & $-0.051$ & $0.213\pm0.064$ & $2.31 \pm0.71$      \\             
{CPD-$59^{\circ}2635$}        & $-6.783\pm0.021$ & $2.582\pm0.019$ & $+0.148$ & $-0.030$ & $0.151\pm0.067$ & $1.68 \pm0.75$      \\                                   
{CPD-$59^{\circ}2641$}        & $-7.015\pm0.017$ & $2.713\pm0.015$ & $-0.084$ & $+0.101$ & $0.131\pm0.062$ & $1.46 \pm0.69$      \\          
{CPD-$59^{\circ}2644$}        & $-6.799\pm0.018$ & $2.541\pm0.017$ & $+0.132$ & $-0.071$ & $0.150\pm0.064$ & $1.67 \pm0.72$      \\    
{HD 303316}                            & $-6.401\pm0.016$ & $2.083\pm0.016$ & $+0.530$ & $-0.529$ & $0.749\pm0.063$ & $8.34 \pm0.70$      \\
{ALS 15210}                             & $-7.226\pm0.016$ & $2.828\pm0.016$ & $-0.295$ & $+0.216$ & $0.366\pm0.063$ & $4.08 \pm0.71$       \\
{HD 93162}                               & $-6.918\pm0.023$ & $2.764\pm0.023$ & $+0.013$ & $+0.152$ & $0.153\pm0.062$ & $1.70 \pm0.69$     \\
{HD 93343}                               & $-6.965\pm0.018$ & $2.311\pm0.016$ & $-0.034$ & $-0.301$  & $0.303\pm0.060$  & $3.38 \pm0.67$      \\
 {\bf  Mean values}                   & ${\bf -6.931 \pm 0.063}$  & ${\bf 2.612 \pm 0.058}$ &       &          & ${\bf 0.311\pm0.044}$ & ${\bf 3.46\pm0.49}$ \\                                                                             
                                                     &                                   &                                  &                             & &               \\                                                                                                                                 
{\bf Bochum 11:}                       &                                   &                                  &                             &  &              \\    
{HD 93576}                              & $-6.183\pm0.020$ & $1.967\pm0.018$ & $+0.130$ & $+0.071$ & $0.148\pm0.042$ & $1.65 \pm0.47$    \\   
{HD 93632}                              & $-6.442\pm0.021$ & $1.923\pm0.021$ & $-0.129$  & $+0.027$ &$0.132\pm0.049$  & $1.47 \pm0.55$     \\
{HD 305539}                            & $-6.363\pm0.014$ & $1.853\pm0.014$ & $-0.050$  & $-0.043$  & $0.066\pm0.041$ & $0.74 \pm0.45$     \\
{HD 305612}                            & $-6.251\pm0.017$ & $1.824\pm0.018$ & $+0.062$ & $-0.072$  & $0.095\pm0.039$ & $1.06 \pm0.44$    \\
{ALS 18083}                             & $-6.325\pm0.027$ & $1.915\pm0.029$ & $-0.012$  & $+0.019$ & $0.022\pm0.044$ & $0.25 \pm0.49$      \\
{\bf Mean values}                     &${\bf -6.313 \pm 0.045}$ & ${\bf 1.896 \pm 0.026}$ &  &                 & ${\bf 0.093\pm0.023}$ & ${\bf 1.04\pm0.25}$   \\     

\enddata

\tablenotetext{a} {Proper motion components (mas/yr) in RA and Dec, in solar frame (columns 2 and 3)
and Carina frame (columns 4 and 5) after subtracting mean values for each cluster (boldface).  
Errors on the mean, $\sigma/\sqrt N$, are derived from sample variance $\sigma$ and number $N$ in 
each cluster.  Column~6 gives total parallax in cluster frame, from 
${\rm PM}_{\rm tot}^2 = {\rm PM}_{\rm RA}^2 + {\rm PM}_{\rm Dec}^2$.   Column 7 gives the star's 
transverse velocity (1 mas/yr is 11.14 \kms\ at 2.35~kpc).}

\end{deluxetable}


 \end{document}